%% file: all_driver.tex
\documentclass[]{jfm}

\usepackage{graphicx}
\begin{document}
\input symbols.tex
\newcommand{\bx}{{\bf x}}
\newcommand{\bv}{{\bf v}}
\newcommand{\bj}{{\bf j}}
\newcommand{\bB}{{\bf B}}
\newcommand{\bF}{{\bf F}}
\newcommand{\bOmega}{{\bf \Omega}}
\newcommand{\bomega}{{\bf \omega}}
\newcommand{\bnabla}{{\bf \nabla}}
\newcommand{\apjl}{{Astrophysical Journal Letters}}
\newcommand{\apj}{{Astrophysical Journal}}
\newcommand{\physrep}{{Physics Reports}}
\newcommand{\mnras}{{Mon. Not. Roy. Ast. Soc.}}
\newcommand{\pre}{{Phys. Rev. E}}
\newcommand{\aap}{{Astronomy \& Astrophysics}}
\title{MHD Dynamos and Turbulence}
\author[S. M. Tobias, F. Cattaneo and S. Boldyrev]
{S\ls T\ls E\ls V\ls 
E\ls N\ns  M.\ns  T\ls O\ls B\ls I\ls A\ls S, F\ls A\ls U\ls S\ls T\ls O \ns C\ls A\ls T\ls T\ls A\ls N\ls E\ls O
\and S\ls T\ls A\ls N\ls I\ls S\ls L\ls A\ls V\ns 
B\ls O\ls  L\ls  D\ls Y\ls R\ls E\ls V\ls
}
%\affiliation{$^1$ Department of Astronomy and Astrophysics and The Computation Institute, University of Chicago,
%Chicago, IL 60637, USA \\[\affilskip]
%$^2$ Department of Applied Mathematics, University of Leeds, Leeds LS2 9JT, UK
%\title[turbulence etc]
%      {}
%\author{S.M.~Tobias, F. Cattaneo \& S. Boldyrev}
\date{29 January 2010}

\pagenumbering{roman}
\maketitle
%\tableofcontents
%\cleardoublepage
%\pagenumbering{arabic}

%\chapter{MHD turbulence}
\author[S.M.~Tobias, F. Cattaneo \& S. Boldyrev]{S.M.~Tobias, F. Cattaneo \& S. Boldyrev}
%\maketitle

%Stuff here
\input introduction.tex

\input dynamo_new.tex

\input meanfield.tex

\input conclusions.tex

%\bibliography{References_all}
%\bibliographystyle{jfm}

\input all_driver_edited.bbl
\end{document}

%% file: symbols.tex
\newcommand{\bfv}{{\bf v}}
\newcommand{\bfu}{{\bf u}}
\newcommand{\bfb}{{\bf b}}
\newcommand{\bfB}{{\bf B}}
\newcommand{\bfJ}{{\bf J}}
\newcommand{\bfF}{{\bf F}}
\newcommand{\bfx}{{\bf x}}
\newcommand{\bff}{{\bf f}}
\newcommand{\bfA}{{\bf A}}
\newcommand{\half}{\frac{1}{2}}

%% file: introduction.tex
%  Steve this is the latest version (10-02-2009)

\section{Introduction}

Magnetic fields are ubiquitous in the universe (\cite{Parker:1979, zelrussok:1983}). Their interaction with an electrically conducting fluid gives rise to a complex system--a %%@
magnetofluid--whose dynamics is quite distinct from that of either a non conducting fluid, or that of a magnetic field in a vacuum (\cite{cowling:1978}). The scales of these interactions %%@
vary in nature from metres to megaparsecs and  in most situations, the dissipative processes occur on small enough scales that the resulting flows are turbulent. 
The purpose of this review is to discuss a small fraction of what is currently known about the properties of these turbulent flows. We refer the reader to several recent reviews for a %%@
broader view of the field (\cite{biskamp:2003, galtier:2008, galtier:2009, lazarian:2006, lazariancho:2005, mullerbusse:2007, kulsrudzweibel:2008}, Bigot {\em et~al.} 2008, Sridhar 2010, Brandenburg \& Nordlund 2010).
The electrically conducting fluid most commonly found in nature is
ionized gas, i.e. a plasma, and its description in terms of all its
fundamental constituents is extremely complex (see e.g. %%@
\cite{kulsrud:2005}). In many circumstances, however, these complexities can be neglected in favour of a simplified description in term of a single fluid interacting with a magnetic field. %%@
Formally, this approach is justifiable when the processes of interest occur on timescales long compared with the light-crossing time, and on spatial scales much larger that any %%@
characteristic plasma length. Despite its simplified nature, the resulting Magneto-Hydrodynamic (MHD) approximation is of great general applicability and can adequately describe many %%@
astrophysical systems ranging from galaxy clusters, to the interstellar medium,  to stellar and planetary interiors, as well as laboratory experiments in liquid metals. The dynamics of turbulent flows under this approximation will form the %%@
basis for our discussion. 

There are many similarities between turbulence in a magnetofluid and in a non-electrically conducting fluid --- hereinafter a regular fluid. The most fundamental is the idea of a turbulent %%@
cascade whereby energy is transferred from a large injection scale by nonlinear interactions through an inertial range to small scales where it is converted into heat by dissipative %%@
processes (see e.g. \cite{davidson:2004}). In analogy with regular turbulence, one of the objectives of theories of MHD turbulence is to predict the general form of the inertial range and %%@
dissipative subrange, and to identify the important processes therein. There are, however, many significant differences. First of all, MHD describes the evolution of two vector fields, the %%@
velocity and the magnetic field, and hence the specification of the state of the system is more involved. 
A further crucial difference concerns the influence of global constraints. In regular turbulence global constraints become progressively weaker at small scales, and in general it is always %%@
possible to find a scale below which the motions become unconstrained. For example, in a rotating fluid scales smaller than the Rossby radius hardly feel the rotation at all;  in a %%@
stratified fluid, motions on a scale much smaller than a representative scale height hardly feel the stratification.
In contrast, there is no scale in MHD turbulence below which the fluid becomes unmagnetized. This has an important consequence: the cascade in regular turbulence approaches an isotropic state at small %%@
scales, whereas the cascade for MHD turbulence becomes progressively more anisotropic at small scales, this being true even if there is no large-scale field.  
Moreover, an important effect in MHD turbulence that has no direct correspondence in regular turbulence is the so-called dynamo process whereby a turbulent electrically conducting fluid %%@
becomes self magnetized. Finally, there are several differences between regular and MHD turbulence that are social in origin. We are constantly immersed in regular turbulence. We have a %%@
direct experience of it in our everyday life. Thus our development of models and theories of regular turbulence is both strongly guided and strongly constrained by experimental data and %%@
intuitions. Not so for MHD turbulence. Even though MHD turbulence is very widespread in the universe, we have practically no direct experience of it in our daily pursuits. Laboratory %%@
experiments can be conducted but they are, in general difficult to perform, and difficult to diagnose (\cite{verhille-etal:2009}). As a result, MHD theorists have enjoyed a greater creative %%@
freedom and have expressed it by generating an impressive array of competing theories; computer simulations being the only obstacle standing in the way of the expression of unbounded %%@
imagination. As a testament to this, we note that whereas the form of the energy spectrum in isotropic homogeneous (regular) turbulence was very much a settled issue by the late forties %%@
(for a historical perspective see \cite{Frisch:1995}), the corresponding problem in MHD turbulence is still very much open to debate today.

\subsection{Formulation and equations}

In this review we shall restrict our discussion to incompressible
MHD. In terms of the fluid velocity this is appropriate for subsonic
flows. In MHD an additional constraint on the magnitude of the
magnetic fluctuations is also required for the applicability of
incompressibility. It is %%@
customary to define the plasma $\beta$ as the ratio of the  gas
pressure to the magnetic pressure --- then  incompressibility is appropriate for situations in which $\beta \gg 1$.  This %%@
condition is satisfied, for instance,  in planetary and stellar interiors, but not in dilute plasmas such as those found in the solar corona and in many fusion devices (\cite{kulsrud:2005}). We note here also that in some specific circumstances the incompressible equations provide a good approximation even if $\beta \ll 1$.
The evolution of the magnetofluid is described by the momentum equation together with the induction equation and the requirements that both the velocity and magnetic fields be solenoidal. In %%@
dimensionless form, these can be written as

\begin{eqnarray}
&
\partial_t \bfv + \bfv \cdot \nabla \bfv =  - \nabla p + \bfJ \times \bfB + Re^{-1} \nabla^2 \bfv + \bff,
&\label{eq_mom}
\\
&
\partial_t \bfB = \nabla \times \left( \bfv \times \bfB \right) + Rm^{-1} \nabla^2 \bfB,
&\label{eq_ind}
\\
&
\nabla\cdot\bfB=\nabla\cdot\bfv=0,
&\label{eq_div}
\end{eqnarray}
where $\bfv$ is the fluid velocity, $\bfB$ is the magnetic field intensity, $p$ the pressure, $\bfJ$ is the cuurent density, and $\bff$ is the total body force, that could include rotation, %%@
buoyancy, etc. Here, $Re=v_o \ell /\nu$ in the momentum equation is the familiar Reynolds number and $Rm$ in the induction equation is the corresponding magnetic Reynolds number defined %%@
analogously by $Rm=v_o \ell /\eta$, where $\eta$ is the magnetic diffusivity; it measures the stength of inductive processes relative to diffusion. As is customary, we have assumed that the %%@
density $\rho_o$ is constant and uniform, and we measure $\bfB$ in units of the equivalent Alfv\'en speed ($B/\sqrt{4 \pi \rho_o}$). 
The ratio of the magnetic to the kinetic Reynolds number is a property of the fluid and is defined as the magnetic Prandtl number $Pm$. In most naturally occurring systems, $Pm$ is either %%@
extremely large, like in the interstellar medium, the intergalactic medium and the solar wind, or extremely small like in the dense plasmas found in stellar interiors or liquid metals. %%@
Interestingly, it is never close to unity except in numerical simulations. This has some important consequences in terms of what we can compute as opposed to what we would like to %%@
understand; we shall return to this issue later. In this review, we shall concentrate on the cases in which $Re$ is large, so that the flows are turbulent, and $Rm$ can be small, as in %%@
some liquid metal experiments, moderate, as in planetary interiors, or large as in most astrophysical situations. 

For an ideal fluid, i.e.\ one in which $Rm$ is infinite, magnetic field lines move with the fluid as if they were frozen in. This is analogous to the behaviour of vortex lines in an inviscid %%@
fluid. Indeed, Alfv\'en's Theorem states that the magnetic flux through  a material surface is conserved in the same way as Kelvin's theorem asserts that the circulation of a material %%@
contour is conserved (\cite{cowling:1978, Moffatt:1978}). In fact, there is a formal analogy between the induction and vorticity equations. It should be noted however that, whereas the %%@
vorticity is the curl of the velocity, no such relationship exists between $\bfv$ and $\bfB$. As a result, arguments based on the formal analogy between the two equations can sometimes be %%@
useful, and sometimes be misleading.  In general, as a rule of thumb, if the magnetic field is weak compared with the velocity it behaves analogously to the vorticity; if it is comparable it %%@
behaves like the velocity. This point will be discussed more fully later. 

As in hydrodynamic turbulence, much insight can be gained from examining conservation laws. In the ideal limit there are three quadratic conserved quantities: the total energy 
\begin{equation}
E = \half \int_V \left( v^2+B^2 \right) \, d^3 x,
\label{def_energy}
\end{equation}
 the cross-helicity 
\begin{equation}
H^c = \int_V \bfv \cdot \bfB \, d^3 x,
\label{def_cross}
\end{equation}
 and the magnetic helicity
\begin{equation}
H = \int_V \bfA \cdot \bfB \, d^3 x,
\label{def_maghel}
\end{equation}
where $\bfA$ is the vector potential satisfying $\bfB = \nabla \times
\bfA$. Here the volume $V$ is either bounded by a material flux
surface, or is all space provided that %%@
the fields decay sufficiently fast at infinity
(\cite{Moffatt:1978})\footnote{Many numerical simulations utilise
  periodic boundary conditions, for which care must be taken in
  defining the magnetic helicity.}. It should be noted that the limit $\bfB \rightarrow 0$ is a delicate one; only the energy survives as a conserved %%@
quantity and a new conserved quantity the kinetic helicity appears. This emphasises the fundamental difference between hydrodynamics and MHD. It can be argued analytically and verified %%@
numerically that, in the presence of small dissipation, energy decays faster than magnetic helicity and cross-helicity (\cite{biskamp:2003}). Therefore, in a turbulent state, energy cascades %%@
toward small scales (analogous to the energy cascade in 3D hydrodynamics), with the  magnetic fluctuations approximately in equipartition with the velocity fluctuations. The magnetic %%@
helicity cascades toward large scales (analogous to the energy cascade in 2D hydrodynamics); the  inverse cascade of magnetic helicity may lead to the formation of large-scale magnetic %%@
fields, which are not in equipartition with the velocity fluctuations (\cite{pouquetetal:1976}).  
The role of cross-helicity is more subtle. Having the dimension of energy, it also cascades towards small scales (\cite{biskamp:2003}). However, cross-helicity dissipation is not %%@
sign-definite: cross-helicity may be either amplified or damped locally. As we shall see, this  leads to local self-organization in the turbulent inertial interval.

We can now discuss the most significant difference between hydrodynamic and MHD turbulence. In hydrodynamic turbulence, mean flows or large-scale eddies advect small-scale fluctuations %%@
without affecting their dynamics (\cite{batchelor:1953}). This is a consequence of the Galilean invariance of the Navier-Stokes equation. The situation is quite the opposite in MHD %%@
turbulence. While the large-scale velocity field can be removed by Galilean transformation, the large-scale magnetic field cannot. Such a large-scale magnetic field could arise owing to a %%@
number of different processes. It could  be either generated by large-scale eddies, as in the interstellar medium, or imposed by external sources, as in the solar wind or plasma fusion %%@
devises. The large-scale magnetic field mediates the energy cascade at all scales in the inertial interval.  As a consequence, as we pointed out earlier, weak small-scale turbulent %%@
fluctuations become anisotropic, since it is much easier to shuffle strong magnetic field lines than to bend them. The smaller the scale, the stronger the anisotropy caused by the %%@
large-scale field. This is in a stark contrast with hydrodynamic turbulence where large-scale anisotropic conditions get ``forgotten'' by smaller eddies, so that the turbulent fluctuations %%@
become isotropic as their scale decreases.

In this review, we shall organise our discussion into two main sections describing the dynamics of turbulence with and without a significant mean field. If the mean field is unimportant then %%@
we neglect it altogether and consider the ``dynamo'' case, whereas if it is important we shall assume that it is strong and shall not concern ourselves about its origin. One should note %%@
that this distinction may depend on scale and the same system may be well described by the ``dynamo'' case at some scales and the ``guide-field'' case at others. Finally, we shall mostly be %%@
concerned with driven turbulence, that eventually evolves to a stationary state. There is a substantial body of literature considering turbulent decay (see e.g. \cite{biskamp:2003}) that %%@
will not be discussed here.%%@

%% file: dynamo_new.tex
%   Figures 
%   Figure 5b   (Carati et al 2006 figure 3)

\section{Dynamo}

%\cite{tdh2007}
%\citep{catttob2009}

In this section we consider the case where there is no externally imposed magnetic field. It is well known that the %%@
turbulent motion of 
an electrically conducting fluid can lead to the amplification of a seed magnetic field (\cite{Parker:1979,Moffatt:1978}). %%@
This generation process is termed dynamo action and can lead
to a substantial level of magnetisation. Two questions naturally arise. Under what conditions does dynamo action take %%@
place and what is the 
final state of the turbulence and magnetic field in the magneto-fluid? We discuss these questions in turn.

\begin{figure}
\begin{minipage}[b]{0.5\linewidth} % A minipage that covers half the page
\centering
\includegraphics[width=0.99\textwidth]{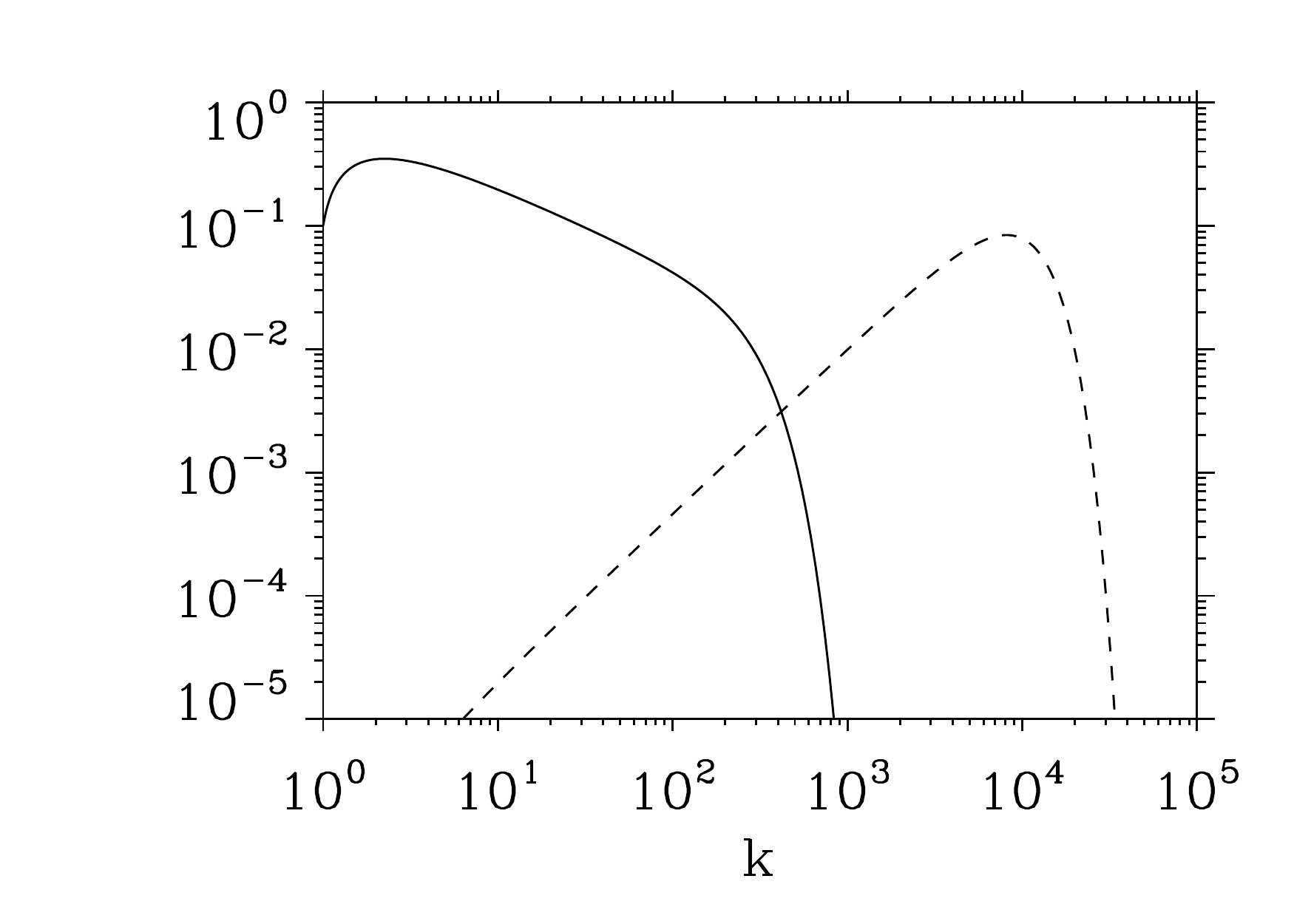}
\end{minipage}
\hspace{0.1cm} % To get a little bit of space between the figures
\begin{minipage}[b]{0.5\linewidth}
\centering
\includegraphics[width=0.99\textwidth]{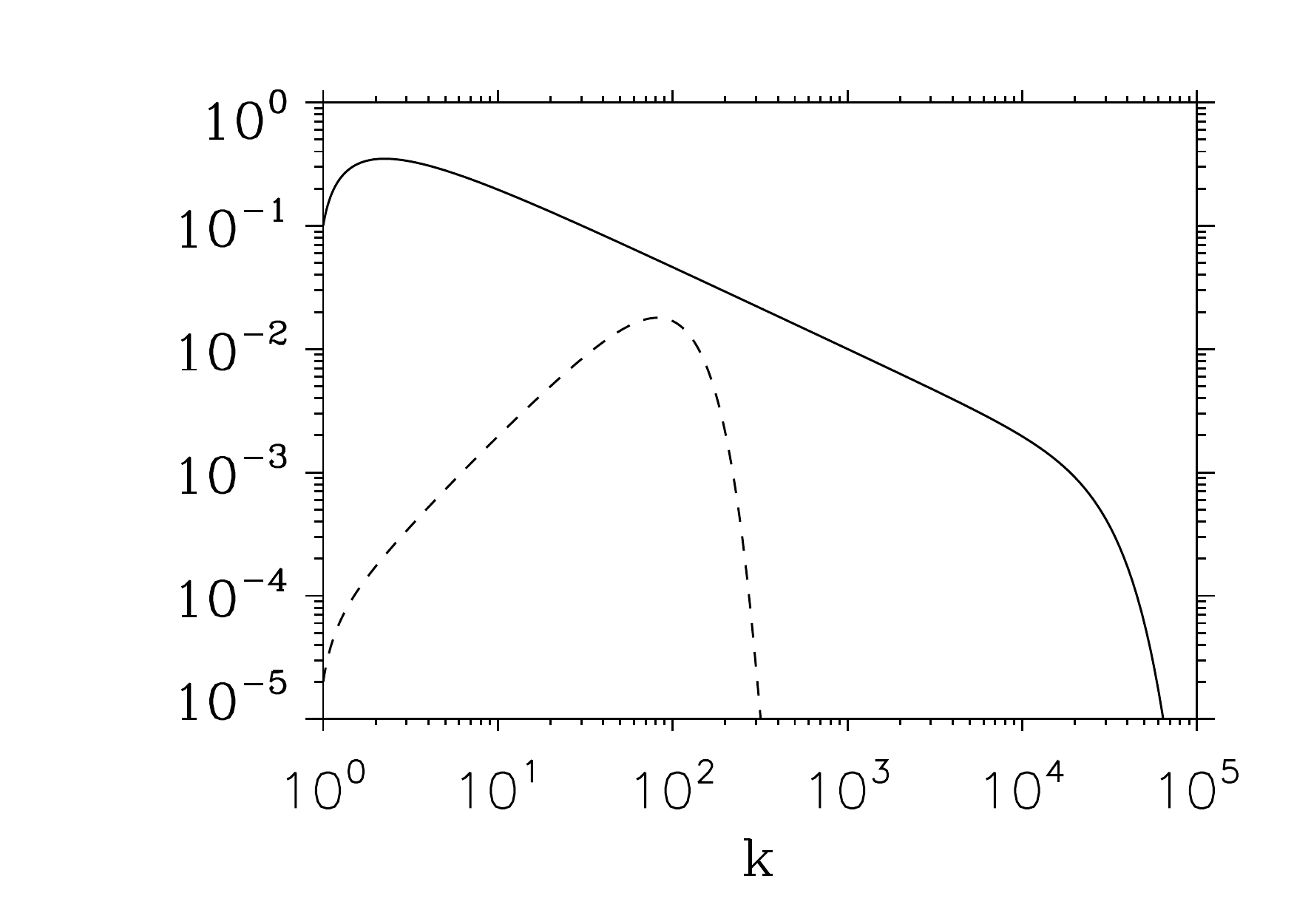}
\end{minipage}
\caption{Possible schematic spectra for kinetic energy (solid line) and magnetic energy (dashed line) for dynamo action %%@
at (a) high and (b) low $Pm$.
When $Pm$ is large the resistive scale for the  magnetic field  is much smaller than a typical eddy and so the flow %%@
appears large-scale and smooth. When $Pm$ is small, the resistive scale lies in the inertial range of the turbulence %%@
where the velocity is rough.}
\label{low_vs_high_pm}
\end{figure}

\subsection{The onset of turbulent dynamo action}

We consider a situation in which an electrically conducting fluid, confined within some bounded region of space, is in a %%@
state of
turbulent motion, driven by  stationary forces. At some time, a very weak magnetic perturbation is introduced into the %%@
fluid. The turbulent motions will stretch the magnetic field and
will therefore typically lead either to its amplification or its decay. In the case of growth the field will eventually %%@
become strong, and the nonlinear Lorentz force term in the momentum
equation will become important. This force will modify the turbulence in such a way that the field no longer grows and %%@
the system will settle down to a stationary state. However, initially, as the initial field is small, the
Lorentz force is negligible and the question of growth versus decay of the field can be addressed by assuming that the %%@
velocity is given and by solving the induction
equation alone. This defines the kinematic dynamo problem.

For a prescribed velocity field, this is a linear problem for the magnetic field $\bB$. There are three cases that are %%@
typically considered within the literature: for a steady velocity
one can seek solutions of the form $\bB = \bB({\bf x}) \exp \sigma t$ and the induction equation becomes a classical %%@
eigenvalue problem for the dynamo growth-rate $\sigma$. The value of $Rm$
for which the real part of the growth-rate ($\Re(\sigma)$) becomes positive is termed the critical magnetic Reynolds number $Rm^{crit}$ for the onset of %%@
dynamo action. Another commonly considered case is one where the velocity is periodic in time, then the induction %%@
equation defines a Floquet problem and solutions consist of a periodic and an exponentially growing part and again one %%@
can define
$Rm^{crit}$ in analogy with the steady case. Finally there is the case in which the velocity is a stationary random %%@
process. Here one utilises a statistical description of the field, and seeks conditions under which the moments of the %%@
probability distribution function for $B = |\bB|$ grow exponentially (\cite{zelruzsok:1990}). In all of these cases %%@
$Rm^{crit}$ corresponds to that value at which the induction processes overcome diffusion.  Na\"ively one would expect %%@
that once $Rm > Rm^{crit}$ dynamo action would be guaranteed, but
this, actually, is not necessarily so. In fact establishing %%@
dynamo
action in the limit $Rm \rightarrow \infty$, the so-called fast-dynamo problem, is  technically very difficult (see for %%@
instance \cite{chilgil:1995}). For instance it has been shown that flows that do not have chaotic streamlines can not be %%@
fast dynamos (\cite{klapper:1995}). This exemplifies the somewhat paradoxical role of diffusion in dynamo action. On the %%@
one hand too much diffusion suppresses dynamo action, on the other hand not enough diffusion also makes dynamo action %%@
impossible. The reason is that reconnection is required in order to
change the magnetic topology to allow for the growth of the %%@
field 
(see e.g. \cite{dormysoward:2007}). In fact, as we shall see, the diffusive scale at which reconnection occurs plays an important %%@
role in determining the properties of dynamo action.

Once a growing solution has been identified, it is of interest to determine properties of the eigenfunction. The %%@
detailed properties depend on the precise form of the velocity, but typically most of the energy of the growing %%@
eigenfunction is concentrated at the reconnection scales. However there is considerable interest, mostly astrophysically %%@
motivated, in cases where a significant fraction of the energy is found on scales larger than a typical scale for the %%@
velocity. This is called the large-scale dynamo problem and is most commonly discussed within the framework of mean %%@
field electrodynamics (\cite{Moffatt:1978,krauraed:1980}). One of the early successes of this kinematic theory was to %%@
establish that a lack of reflectional symmetry of the underlying flow is a necessary condition for the generation of %%@
large-scale fields. There is a substantial body of literature that discusses the large-scale dynamo problem. It is fair %%@
to say that currently there is considerable controversy as to whether mean-field electrodynamics can be applied in cases %%@
where $Rm$ is large (see e.g. \cite{bransub:2005,bolcatt:2004,hugcatt:2009}). In this present review we shall not discuss at %%@
length the large-scale problem, but instead focus on the amplification and saturation of magnetic fields of any scale.

Although anti-dynamo theorems exclude the possibility of dynamo action for certain flows and magnetic fields that %%@
possess too much symmetry, for example two-dimensional flows or axisymmetric fields (\cite{cowling:1933,Zeldovich:1956}), %%@
it is
well established that most sufficiently complicated laminar flows do lead to growing fields at sufficiently high $Rm$. %%@
However here we are interested in the corresponding question for turbulent flows. In order to address this issue we need %%@
to discuss two distinct possibilities corresponding to the cases of small and large magnetic Prandtl number $Pm$.

\subsection{High $Pm$ versus low $Pm$, smooth versus rough}

For the purposes of this discussion, we shall assume that the fluid Reynolds number is high so that the flow is turbulent %%@
with a well-defined inertial range that extends from the integral scale $l_0$ to the dissipative scale $l_\nu$, and a %%@
dissipative sub-range for the scales $l < l_\nu$.
It is useful to define the second order structure function $\Delta_2(r) = \langle |(\bv(\bx,t) - \bv(\bx+{\bf %%@
r},t)).{\bf e}_r|^2  \rangle$, where $r = |{\bf r}|$ and ${\bf e}_r={\bf r}/r$, where we have assumed homogeneity and %%@
isotropy. We can then characterise the inertial and dissipative ranges by the scaling exponents of $\Delta_2(r)$ with %%@
$\Delta_2(r) \sim r^{2 \alpha}$, where $\alpha$ is the roughness exponent. In the dissipative sub-range we expect the %%@
velocity to be a smooth function of position and therefore $\alpha=1$, whilst for the inertial range the velocity %%@
fluctuates rapidly in space, i.e. it is {\it rough} and  $\alpha < 1$;
for example,  for Kolmogorov turbulence $\alpha = 1/3$. %%@
The inertial range can also be characterised by the slope of the energy spectrum $E_k \sim k^{-p}$ where  $p$ is related %%@
to the  roughness exponent by $p = 2 \alpha+1$. The dissipative scale is also related to $\alpha$ and given by 
$l_\nu = l_0 Re^{-1/1+ \alpha}$.

For this given velocity field the scale at which magnetic dissipation becomes important and reconnection occurs can be %%@
analogously defined by $l_\eta = l_0 Rm^{-1/1+ \alpha}$. The ratio
of the two dissipative scales is given by $l_\nu / l_\eta = Pm^{1/1+ \alpha}$. Clearly, irrespective of the value of %%@
$\alpha$, if $Pm \gg 1$ then the resistive scale $l_\eta$ is much smaller than the viscous scale $l_\nu$, and therefore %%@
reconnection occurs in the dissipative sub-range where the velocity is spatially smooth, as in %%@
Figure~(\ref{low_vs_high_pm}a). By contrast, if $Pm \ll 1$ the dissipative scale is much greater than the viscous scale %%@
and therefore reconnection occurs in the inertial range where the velocity is rough and therefore fluctuates rapidly, as %%@
shown in Figure~(\ref{low_vs_high_pm}b). Therefore $Pm=1$ defines the boundary between dynamo action driven by rough and %%@
smooth velocities.

It will become apparent that in general it is harder to drive a dynamo with a rough velocity than with a smooth %%@
velocity. We now examine some specific examples.

\subsection{Random dynamos - the Kazantsev formulation}

\begin{figure}
\begin{minipage}[b]{0.5\linewidth} % A minipage that covers half the page
\begin{center}
%\centering
\includegraphics[width=0.99\textwidth]{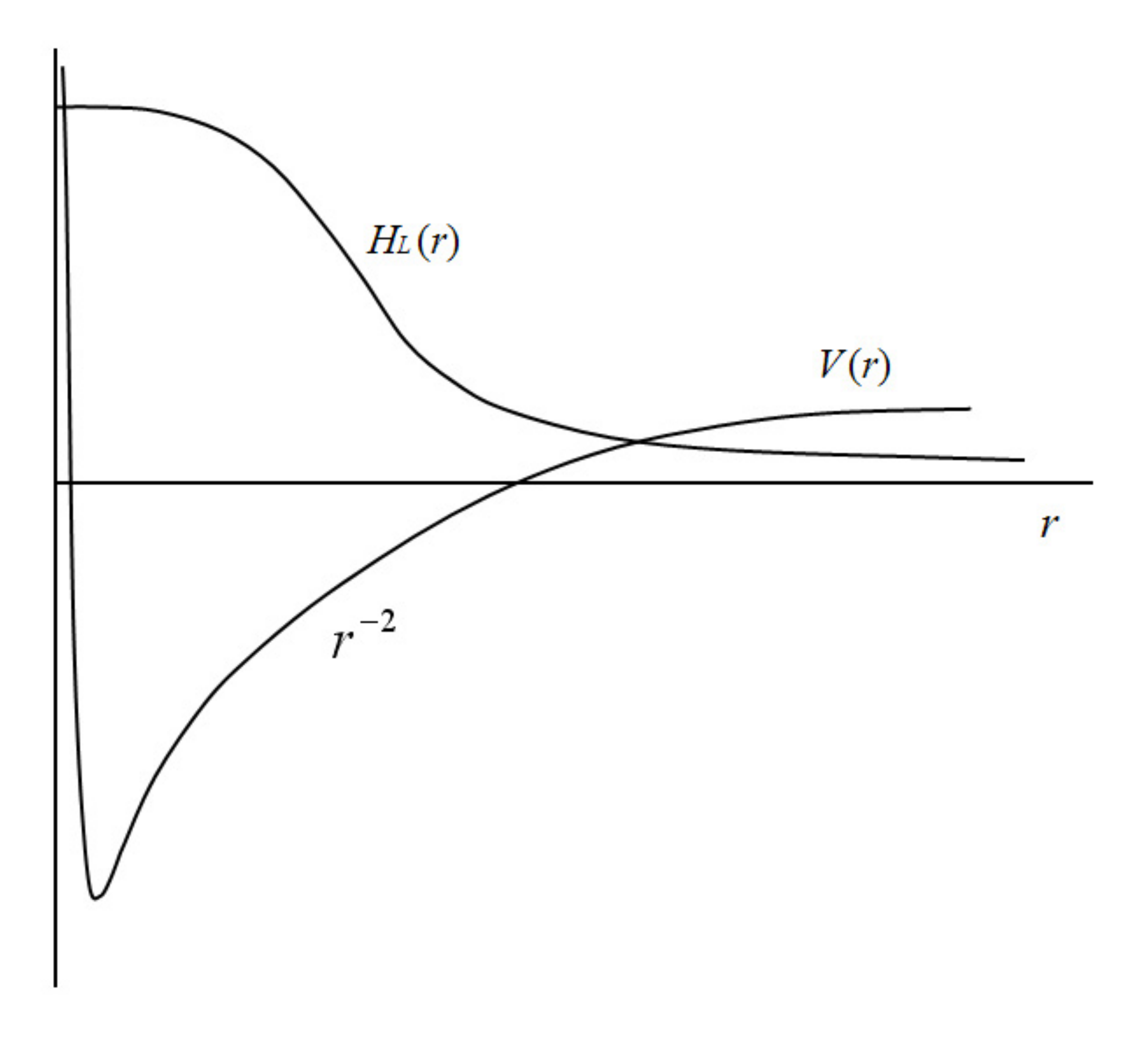}
\end{center}
\end{minipage}
\hspace{0.1cm} % To get a little bit of space between the figures
\begin{minipage}[b]{0.5\linewidth}
%\centering
\begin{center}
\includegraphics[width=0.99\textwidth]{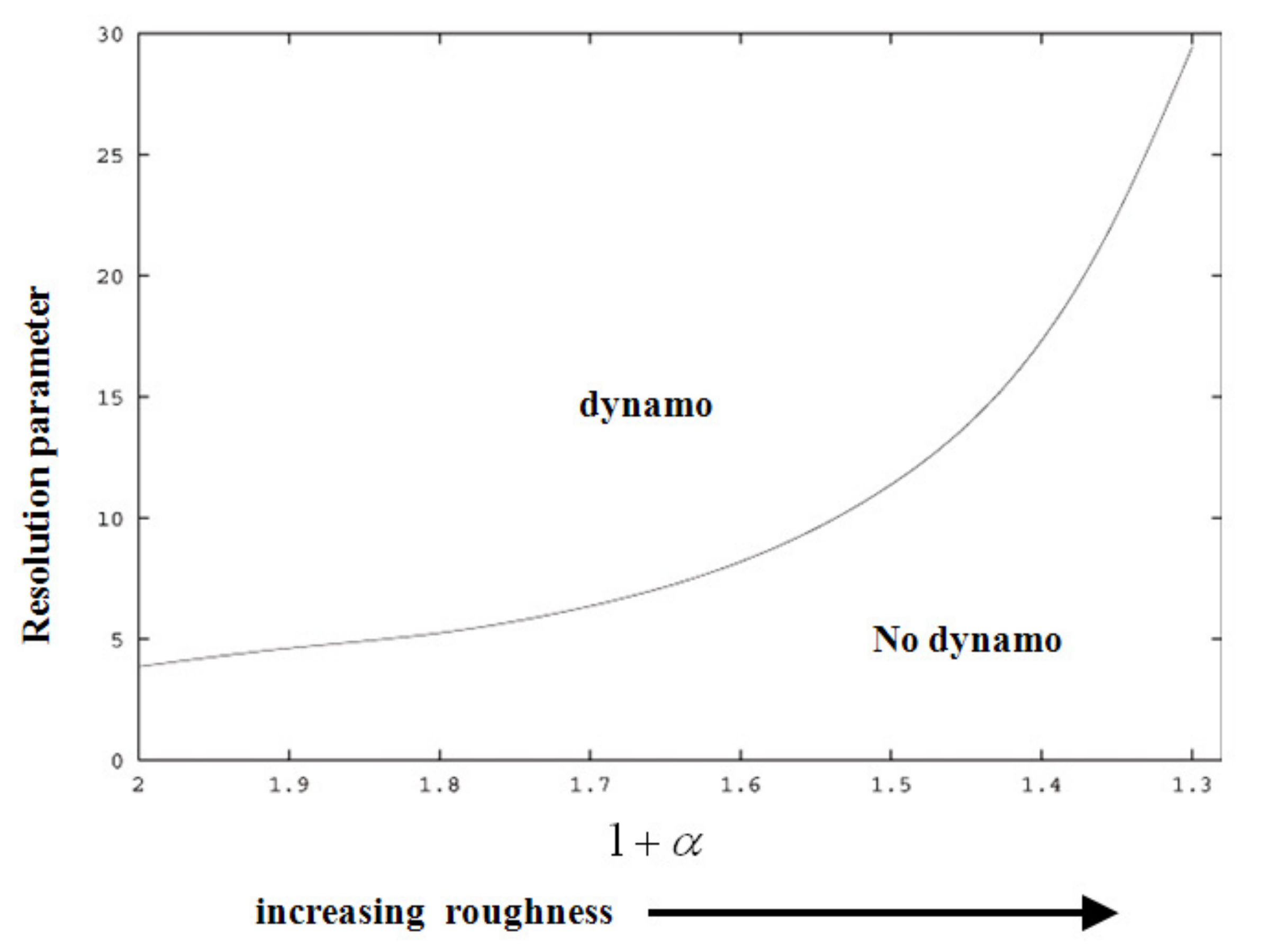}
\end{center}
\end{minipage}
\caption{(a) Schematic representation of the potential $V(r)$ and the spatial part of the longitudinal magnetic correlator
$H_L(r)$ in the Kazantsev model. The potential has a $1/r^2$ behaviour in the inertial range. Its overall height is
 determined by the velocity roughness, and its large $r$ behaviour by the boundary conditions. The magnetic correlator is 
 peaked at small scales and decays exponentially for large values of $r$.
 (b) Critical value of the size parameter $L$ for dynamo action in the Kazantsev model as a function of the 
 velocity roughness $1+\alpha$. As the velocity becomes rougher the effort required to drive a dynamo increases}
% \caption{.}
\label{fig_2}
\end{figure}

%\begin{figure}
%\begin{minipage}[b]{0.5\linewidth} % A minipage that covers half the page
%\centering
%\includegraphics[width=6cm]{bild1.eps}
%\caption{En liten bild}
%\end{minipage}
%\hspace{0.5cm} % To get a little bit of space between the figures
%\begin{minipage}[b]{0.5\linewidth}
%\centering
%\includegraphics[width=6cm]{bild2.eps}
%\caption{En liten bild till}
%\end{minipage}
%\end{figure}

% \begin{figure}
% \centering
%\includegraphics[width=1.0\textwidth]{nj243354fig1.pdf}
% \includegraphics[width=1.0\textwidth]{Figure_2.pdf}
% \caption{Schematic representation of the potential $V(r)$ and the spatial part of the longitudinal magnetic correlator %%@
% $H_L(r)$ in the Kazantsev model. The potential has a $1/r^2$ behaviour in the inertial range. Its overall height is %%@
% determined by the velocity roughness, and its large $r$ behaviour by the boundary conditions. The magnetic correlator is %%@
% peaked at small scales and decays exponentially for large values of $r$.}
% \label{f2}
% \end{figure}

% \begin{figure}
% \centering
%\includegraphics[width=1.0\textwidth]{nj243354fig1.pdf}
% \includegraphics[width=1.0\textwidth]{Figure_3.pdf}
% \caption{Critical value of the size parameter $L$} for dynamo action in the Kazantsev model as a function of the %%@
% velocity roughness $1+\alpha$. As the velocity becomes rougher the effort required to drive a dynamo increases.
% \label{f3}
% \end{figure}

We now look at the simplest model of (kinematic) dynamo action driven by a random flow, the so-called %%@
Kazantsev model (Kazantsev 1968). This is the only known solvable model for dynamo action and, %%@
despite its simplicity, it captures many of the relevant features of turbulent field generation. 

The model is based on a prescribed Gaussian, delta correlated 
(in time) velocity field, with zero mean and covariance
\begin{eqnarray} 
\langle v_i({\bf x+r}, t)v_j({\bf x}, \tau) \rangle
=\kappa_{ij}({\bf x},{\bf r})\delta(t-\tau).
\label{correlator}
\end{eqnarray}

Isotropy and homogeneity imply that the velocity correlation 
function has the form
\begin{eqnarray}
\kappa_{ij}({\bf r})=\kappa_N(r)\left(\delta_{ij}-\frac{r_ir_j}{r^2} 
\right)+\kappa_L(r)\frac{r_ir_j}{r^2}.
\label{kappa}
\end{eqnarray}
Further, incompressibility gives
$\kappa_N=\kappa_L+(r\kappa_L^\prime)/2$, where the primes denote differentiation with respect to $r$, and  now the
velocity statistics can be characterized by the single scalar function 
$\kappa_L(r)$. 
A corresponding expression for the magnetic correlator can be defined by 
\begin{eqnarray}
\langle B_i({\bf x+r}, t)B_j({\bf x}, t) \rangle=H_{ij}({\bf x},{\bf r}, t),
\label{bcorrelator}
\end{eqnarray}
where, analogously to (\ref{kappa}), the $H_{ij}$ function can be 
represented as
\begin{eqnarray}
H_{ij}=H_N(r,t)\left(\delta_{ij}-\frac{r_ir_j}{r^2} 
\right)+H_L(r,t)\frac{r_ir_j}{r^2}.
\label{hdef}
\end{eqnarray}
Similarly, the 
condition $\nabla\cdot {\bf B}=0$ gives $H_N=H_L+(rH_L^\prime)/2$.

It now remains to determine the equation governing the evolution of $H_L(r,t)$ in terms of the input function
$\kappa_L(r)$. This equation is derived by differentiating equation~(\ref{bcorrelator}) with respect to time, using the %%@
dimensional version of the induction equation and expressions
(\ref{correlator}), (\ref{kappa}) and (\ref{hdef}). Straightforward manipulation yields
\begin{eqnarray}
{\partial_t H}_L=\kappa H_L^{\prime \prime} + \left(\frac{4}{r}\kappa+\kappa' \right)H_L^{\prime}
+\left( \kappa''+\frac{4}{r}\kappa' \right)H_L,
\label{kequation}
\end{eqnarray}
where the 
the `renormalized' velocity correlation 
function $\kappa(r)=2\eta+\kappa_L(0)-\kappa_L(r)$ has been introduced, and $\eta$ is the magnetic diffusivity.

Equation~(\ref{kequation}) was originally derived 
by  Kazantsev (1968), and can be rewritten in a related form 
that formally coincides with the Schr\"odinger equation in imaginary time. 
This is achieved by effecting the change of variable, 
$H_L=\psi(r,t) r^{-2}\kappa(r)^{-1/2}$, to obtain
\begin{eqnarray}
\partial_t \psi=\kappa(r)\psi''-V(r)\psi,
\label{sequation}
\end{eqnarray}
which describes the wave function of a quantum particle with variable 
mass, $m(r)=1/[2\kappa(r)]$, in a one-dimensional potential~($r>0$):
\begin{eqnarray}
V(r)=\frac{2}{r^2}\kappa(r)-\frac{1}{2}\kappa''(r)-\frac{2}{r}\kappa'(r)-
\frac{(\kappa'(r))^2}{4\kappa(r)}.
\end{eqnarray}

Different authors have investigated the solutions of equation~(\ref{sequation}) for various choices of $\kappa(r)$ (see %%@
\cite{zelruzsok:1990, arphorvai:2007,chertetal:1999}).
Here we restrict attention to the two extreme cases in which $l_\eta$ is in the deep dissipative subrange ($Pm \gg 1$) %%@
and the velocity is smooth, and the case where $l_\eta$ is in the inertial range ($Pm \ll 1$) and the velocity is rough, %%@
so that $\alpha < 1$. For the smooth case, exponentially growing solutions of equation~(\ref{kequation}) can be found  %%@
and the magnetic energy spectrum $E_M$ is peaked at the dissipative scale. The spectrum in the range $1/l_\eta < k < %%@
1/l_\nu$ has a power law behaviour with $E_M \sim k^{3/2}$, irrespective of the  spectral index for the velocity in the %%@
inertial range (\cite{kuland:1992}). This regime for a smooth velocity
is also referred to as the Batchelor regime, since this is the regime
studied by Batchelor (1959a) for passive-scalar advection.%%@

For the rough case $\kappa(r) \sim r^{1+\alpha}$ in the inertial range. This scaling follows from the fact that Eq.~(\ref{sequation}) depends only on the time-integral of the velocity correlation function (\ref{kappa}), that is, on the turbulent diffusivity. When matching the Kazantsev model with a realistic velocity field, one therefore needs to match the turbulent diffusivities. In the Kazantsev model the diffusivity is given by $\kappa(r)$, while for a  realistic velocity field it is estimated as $\Delta_2(r)\tau(r)$, where $\tau(r)\sim r/[\Delta_2(r)]^{1/2}$ is the typical velocity decorrelation time. This leads to the inertial-interval scaling $\kappa(r)\sim \Delta_2(r)\tau(r)\sim r^{1+\alpha}$ presented 
above.  The 
Schr\"odinger equation~(\ref{sequation}) therefore has the effective 
potential $U_{eff}(r)=V(r)/\kappa(r)=A(\alpha)/r^2$ in the inertial range,  
where $A(\alpha)=2-3(1+\alpha)/2-3(1+\alpha)^2/4 $. At small scales ($l < l_\eta$) this effective potential is regularised %%@
by magnetic diffusion as shown in Figure~(\ref{fig_2}a). Growing dynamo solutions
correspond to bound states for the wave-function $\psi$, which are guaranteed to exist if $A(\alpha) < -1/4$. Since this %%@
is always the case for $0 < \alpha < 1$, this demonstrates that dynamo action is always possible even in the case of a rough %%@
velocity (\cite{bolcatt:2004}). The corresponding wave-function will be concentrated around the minimum of the potential %%@
at $r \sim l_\eta$ and it will decay exponentially for $r > l_\eta$ (see Figure~\ref{fig_2}). We stress at this point that %%@
the effective potential always remains $1/r^2$ in the inertial range with its depth decreasing as the roughness %%@
increases ($\alpha \rightarrow 0$). This justifies our previous statement that it is harder to drive dynamo action the %%@
rougher the velocity.
An asymptotic analysis of the solution to equation~(\ref{sequation}) demonstrates that the growth-time, $\tau$, is of %%@
the order of the turnover time of the eddies at the diffusive scale ($l_\eta$). This makes good physical sense since %%@
these are the eddies that have the largest shear rate. Furthermore the spatial part of the wave-function for large $r %%@
\gg l_0$ decays exponentially as $\exp \left[-2 \tau^{-1/2}r^{(\alpha-1)/2}/(1-\alpha)\right]$. The presence of the factor $1-\alpha$ %%@
in the denominator implies that if $\alpha$ is close to unity (smooth velocities) the wave-function is localised close %%@
to the resistive scale. On the other hand, if $\alpha$ is close to zero (rough velocity) the wavefunction is more spread %%@
out. This can be used to determine the requirements for the onset of dynamo action, which can be expressed either in %%@
terms of a critical magnetic Reynolds number or in terms of a ``size parameter" $L$. The latter is a more useful measure, %%@
since it relates directly to the effort, computational or experimental, that is needed to achieve dynamo action.
Consider equation~(\ref{sequation}) as a two-point boundary value problem in a finite domain of size $l_0$ (see e.g. %%@
\cite{bolcatt:2004}), with boundary conditions $\psi(0) = \psi(l_0) = 0$. As the velocity becomes rougher the %%@
wave-function spreads out and a  larger domain (i.e.\ larger $l_0$) is required to contain the wave-function. %%@
Figure~(\ref{fig_2}(b)) shows the minimum value of $l_0$ measured in units of $l_\eta$ (which is the size parameter $L$) for %%@
which a growing solution can be found as a function of $1+\alpha$. Clearly there is a dramatic increase in  $L$ as the %%@
velocity becomes rougher (1+$\alpha$ gets smaller). Hence the effort required to drive a dynamo also increases. \footnote{An earlier attempt to solve the Kazantsev model for small $Pm$ was made in (Rogachevskii \& Kleeorin 1997). However, this paper employed an incorrect asymptotic matching procedure for deriving the dynamo threshold and the dynamo growth rate. The correct analysis was presented in \cite{bolcatt:2004}.}

We now return to the question posed at the start of this section, what is the effect of changing $Pm$ for the onset of %%@
dynamo action in a random flow?  At high $Pm$ the dynamo operates in the dissipative sub-range where the velocity is smooth and so the %%@
effort necessary to drive a dynamo is modest. This state of affairs continues until $Pm$ decreases through unity. At %%@
that point, the viscous scale becomes smaller than the resistive scale and the dynamo begins to operate in the inertial %%@
range, where the velocity is rough, with a corresponding increase in the effort required to sustain dynamo action. Once %%@
the dynamo is operating in the inertial range, further decreases in $Pm$ do not make any difference to the effort %%@
required. In terms of critical $Rm$ as a function of increasing $Pm$ the curve takes the form of a low plateau and a high %%@
plateau joined by a sharp increase around $Pm=1$ --- see for example the curves in Figure~3. On the other hand, if a sequence of calculations is carried out at %%@
fixed $Rm$ and increasing $Re$ a sharp drop in the growth-rate $\sigma$ will be observed when $Re \approx Rm$ %%@
(\cite{bolcatt:2004}). If the initial $Rm$ is moderate, this could lead to the loss of dynamo action at some $Re$, which %%@
could be misinterpreted as a critical value of $Pm$ at which dynamo action becomes impossible (\cite{christensenetal:1999, %%@
yousefetal:2003,
schekochihietal:2004c,schekochihinetal:2005c}). We shall return to this theme later.

The Kazantsev model is, of course, based on a number of somewhat restrictive assumptions, one of which is that the flow considered is completely random and has no coherent part. We discuss in the next section the role of coherent structure in dynamo problems. Within the statistical framework there has been substantial %%@
effort to extend the basic model to more general cases. One common criticism is that delta-correlated velocities are %%@
artificial, with real turbulence having correlation times of order the turnover time. However, one should remember that %%@
in most turbulent flows the turnover time decreases as one goes to small scales. Near onset the dynamo growth time can %%@
be very large compared with the turnover time --- and hence the correlation time --- of the small eddies that typically %%@
participate in the dynamo process. Noting that the growing magnetic fluctuations are   predominantly concentrated at the resistive scales, where the relative motion of magnetic-field lines is affected by magnetic diffusion and, therefore, these lines do not separate with the eddy turnover rate, one expects that the assumption of zero correlation time is not as restrictive as may first %%@
appear. 
Once this is realised, it is to be expected that the dynamo behaviour near onset should be similar for cases with short %%@
but finite correlation time to that with zero correlation time, and indeed this is confirmed by models in which the %%@
correlation time of the turbulence is finite 
(\cite{vainkit:1986,kleerogsok:2002}). A second possible extension is to flows that lack reflectional symmetry. In this %%@
case  the velocity and magnetic  correlators are each defined by two functions, one as before related to the energy %%@
density (either kinetic or magnetic), the other related to the  helicity (either kinetic or magnetic) %%@
(\cite{vainkit:1986}). In this case similar analysis leads to the derivation of a pair of coupled Schroedinger-like %%@
equations for the two parts of the magnetic correlator (\cite{vainkit:1986,bergerros:1995,kimhug:1997}). Although the %%@
analysis is now more involved, it is possible to show that the system remains self-adjoint (\cite{bolcattros:2005}). It %%@
can also be shown that for sufficient kinetic helicity  there is the possibility of extended states ---  these are %%@
states that do not decay exponentially at infinity and can be interpreted as large-scale dynamo solutions, in direct analogy with the solutions of the well-known alpha-dynamo model by \cite{steenbeck66}. However, at %%@
large $Rm$ the largest growth-rates remain associated with the localised bound states, so that the overall dynamo %%@
growth-rate remains controlled by small-scale dynamo action %%@
(\cite{bolcattros:2005,malbol:2007,malbol:2008a,malbol:2008b}). Anisotropic versions of the Kazantsev model have also %%@
been constructed by Schekochihin {\it et al.} (2004b).

\begin{figure}
\centering
\includegraphics[width=1.0\textwidth]{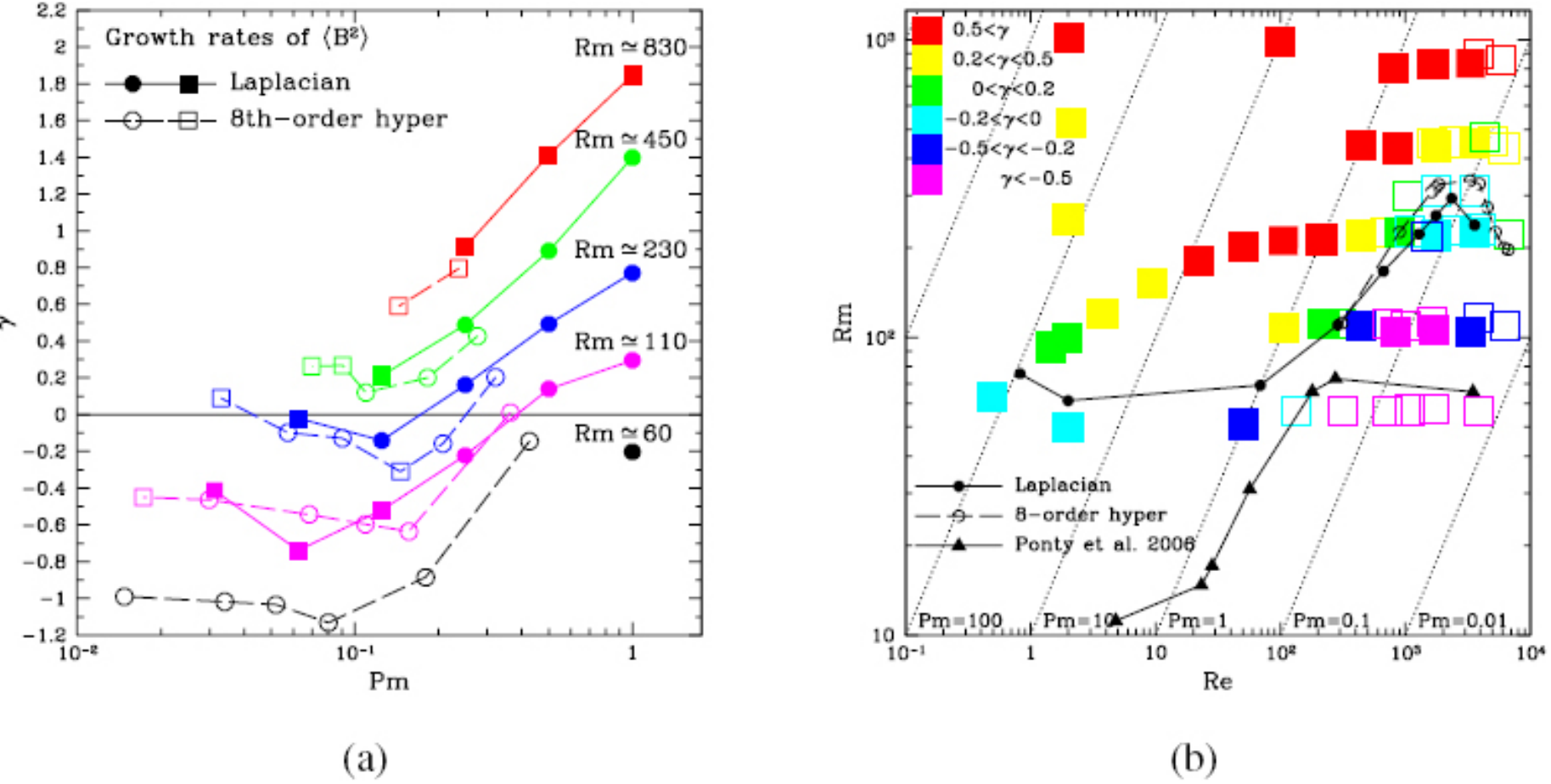}
\caption{Onset of dynamo action at moderate $Pm$, from Schekochihin {\it et al.} (2007). (a) Growth/decay rate of 
magnetic energy as a function of Pm for five values of Rm. (b) Growth/decay rates in the parameter space (Re, Rm). Also %%@
shown are the interpolated stability curves $Rm^{crit}$ as a function of $Re$ based
on the Laplacian and hyperviscous runs are shown separately. For an in-depth discussion of the reason for the apparent
discrepancy between the curves with and without a mean flow see Schekochihin {\it et al.} (2007)}
\label{crit_rm_low_pm}
\end{figure}

Finally there are extensions of the model that take into account departures from Gaussianity. Again on physical grounds %%@
one does not expect this to introduce fundamental changes from the Gaussian case. The reason is similar to the one given %%@
above; near marginality the growth time is long compared with the eddy turnover time and so the dynamo process feels the %%@
sum of many uncorrelated events, which approaches a Gaussian even if the individual events themselves are not. This situation %%@
arises in numerical simulations of dynamo action based on the full solution of the induction equation for velocities %%@
derived from solving the Navier-Stokes equations. Many such efforts are summarised in figure~(\ref{crit_rm_low_pm}), which shows the critical %%@
magnetic Reynolds number as a function of Reynolds number for a collection of such calculations (\cite{scheketal:2007}). %%@
It is clear that at large and moderate $Pm$ these results are consistent with the predictions of the Kazantsev model %%@
above --- the plateau at high $Pm$ is succeeded by a jump in the critical $Rm$ when $Pm$ approaches unity --- such a curve is visible in Figure~\ref{crit_rm_low_pm}. Even the size %%@
of the jump is consistent with the analysis above. However, since the numerical cost of resolving the very thin viscous %%@
boundary layers rapidly becomes prohibitive at small $Pm$, this  regime is not really accessible to direct numerical %%@
simulation (DNS).
One scheme to alleviate this problem is to use large eddy simulations (LES) to generate the velocity %%@
(\cite{pontyetal:2007}). We would advise great caution when using this approach. As the discussion above shows, small %%@
changes in the roughness exponent can lead to huge changes in the critical $Rm$ or alternatively the dynamo growth-rate. %%@
In the case of LES, one would need to be able to control very delicately how the velocity roughness is related to the %%@
smoothing scale. We also note that the jump in critical $Rm$ is captured by shell models of dynamo action (see e.g. %%@
\cite{fricksok:1998}) for which it is possible to achieve a large separation of the viscous and resistive dissipation %%@
scales. We are not certain what, if anything, to make of this.

One of the interesting features of the Kazantsev model is that {\it for random flows} the only thing that matters for %%@
the onset of dynamo action is the roughness exponent of the spectrum in the neighbourhood of the dissipative scale. %%@
Therefore, within this framework, features associated with high-order moments, or large-scale boundary conditions do not %%@
matter. 
Thus from the point of view of dynamo onset a velocity derived from solving the Navier-Stokes equations, with random %%@
forcing should yield qualitatively similar results to a synthesised random velocity with the same spectrum.
So far this seems to be consistent with the results of numerical simulations. However, for flows with a substantial non-random %%@
component, outside of the range of validity of the Kazantsev model it may be that characteristics other than the %%@
spectral slope of the velocity do play a key role in determining dynamo onset, and it is this possibility that we %%@
address in the next section. The results of the next section should therefore act as a caveat when considering the applicability of
the results of this section.

\subsection{Coherent Structure dynamos}

\begin{figure}
\centering
\includegraphics[width=1.0\textwidth]{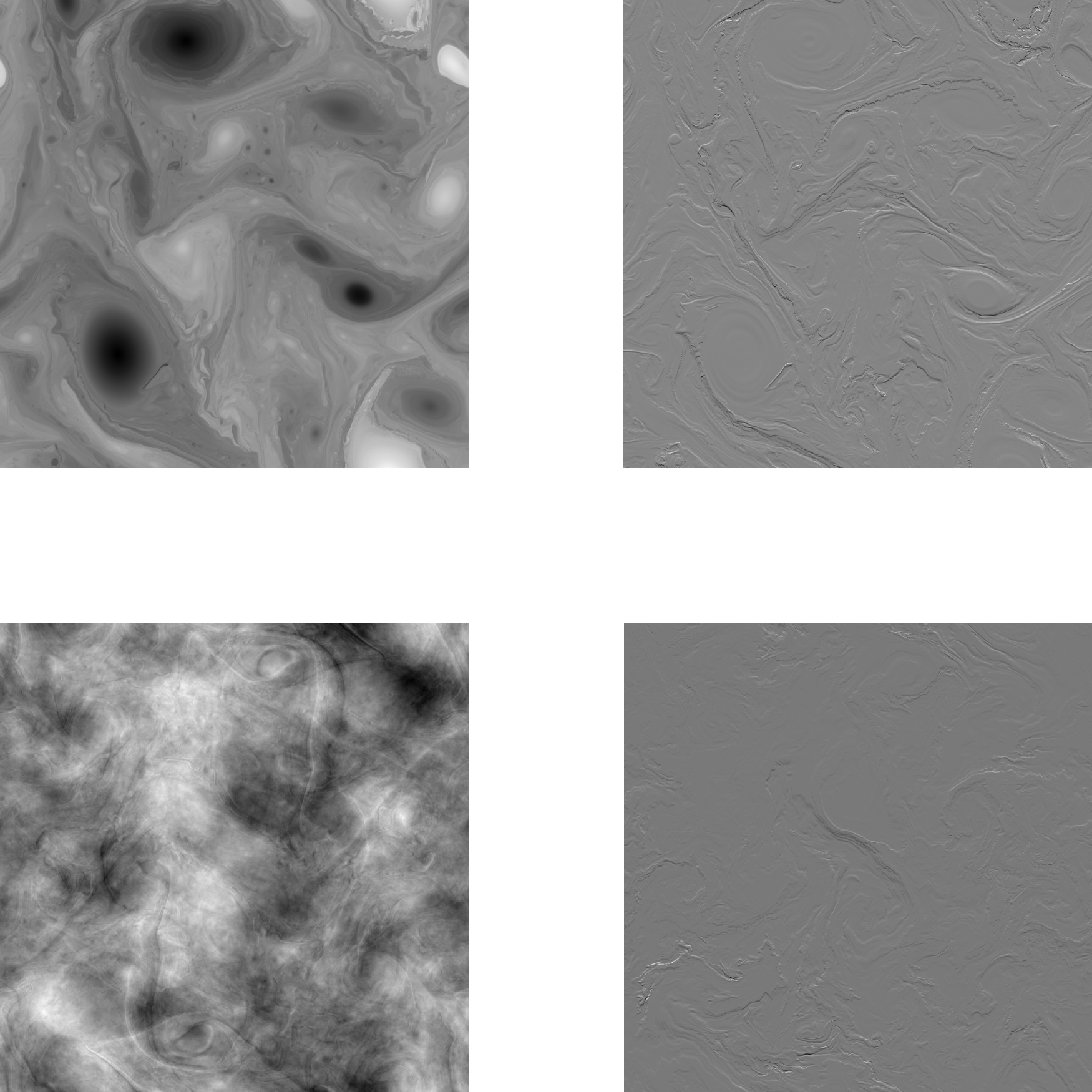}
\caption{Dynamos with and without coherent structures. Out-of-plane velocity (left) and $B_x$ (right) for flows with %%@
coherent structures (top) and
purely random structure (bottom) --- after Tobias \& Cattaneo (2008b). The velocity with coherent structures is a better %%@
dynamo and generates more filamentary magnetic fields.}
\label{coh_stru}
\end{figure}

We now turn to the case where the flow consists of two components, a random component as described above and a more %%@
organised component whose coherence time is long compared with the turnover time (see e.g. \cite{Mininni-etal:2005}). There are many examples of naturally %%@
occuring flows that can be characterised in this way; for example, flows in an accretion disks that consists of %%@
small-scale turbulence and coherent long-lived vortices (see e.g. \cite{Bracco:1999,GodLiv:2000,BalmKor:2001}, Taylor %%@
columns in rapidly rotating turbulent planetary interiors (see e.g. \cite{busse:2002}) or the large-scale flows driven %%@
by propellors in  laboratory experiments (see e.g. \cite{monchauxetal:2007}). In all these cases the turbulence arises %%@
because the Reynolds number is huge, whilst the coherent part is associated with some constraint, such as rotation, %%@
stratification or large-scale forcing.  The natural question then is what determines the dynamo growth-rate? 

In the last section we gained some understanding for the case of entirely random flows --- we know that if the dynamo %%@
were entirely random then the dynamo growth-rate would largely be determined by the spectral index of the flow at the %%@
dissipative scale. Conversely, if the flow were laminar again we could appeal to laminar dynamo theory --- but would not %%@
try to characterise the flow in terms of spectra.
Here we need to consider the case where the flow is the sum of the two components, bearing in mind that the dynamo of %%@
the sum is not the sum of the dynamos (see e.g. \cite{catttob:2005}). We address these issues by examining a specific case. We consider a flow that has %%@
both long-lived vortices and a random component with a well-defined spectrum. For those who are interested, such flows %%@
arise naturally, for instance, as solutions to the active scalar equations (\cite{tobcatt:2008b}). In tandem we also %%@
generate a second flow with the same spectrum but no coherent structures obtained by randomising the phases. Figure~(\ref{coh_stru}) %%@
compares the out-of plane velocity for the two flows. We consider the dynamo properties for large $Rm$ so we anticipate %%@
that dynamo action would occur for either flow. Therefore, if the considerations of the previous section were to apply, %%@
these two flows would exhibit very similar dynamo properties. However, this is actually not so --- the flow with the %%@
coherent structures is a more efficient dynamo in the sense that at the same $Rm$ the dynamo growth-rate is higher. %%@
Moreover the structure of the resulting magnetic field (shown in Figure~\ref{coh_stru}) indicates that the coherent %%@
structures play the major role in field generation. 

Given the above considerations, it is reasonable to ask what controls dynamo action in a flow with a superposition of %%@
coherent structures with different spatial scales and different turnover times? This is equivalent to a turbulent %%@
cascade, but here each of the eddies are long-lived with a coherence time greater than their local turnover time. For %%@
the general case this is a very difficult question to answer. However, there is a case in which some  statements can be %%@
made --- namely when each of the dynamo eddies are `quick dynamos'. A `quick dynamo' (as defined by %%@
\cite{tobcatt:2008a}) is one that reaches a neighbourhood of its maximal growth-rate quickly as a function of $Rm$. In %%@
that case it can be shown that the dynamo is driven by the coherent eddy which has the shortest turnover time $\tau$ and %%@
has $Rm \geq {\cal O}(1)$. Since both the turnover time of the eddy and the local $Rm$ are functions of the spatial %%@
scale of the eddy and the slope of the spectrum, the location in the spectrum of the eddy responsible for dynamo action %%@
also depends on the spectral slope as well as the magnetic Reynolds number on the integral scale (\cite{tobcatt:2008a}).

Some of these ideas become particularly germane in explaining the behaviour of dynamos in liquid metal experiments. The %%@
typical laboratory dynamo device consists of a confining vessel and propellors, designed to drive a large-scale flow %%@
with desirable laminar dynamo properties; i.e.\ with low $Rm^{crit}$. Because the magnetic Prandtl number of liquid %%@
metals is tiny ${\cal O} (10^{-6})$  the corresponding Reynolds numbers are huge. Thus the actual flow consists of the %%@
large-scale flow planned by the designers plus turbulent fluctuations that can be comparable in magnitude to the mean %%@
flow. Invariably, it has been found that the actual critical magnetic Reynolds number for the mean flow and turbulence %%@
is significantly higher than that envisaged for the laminar flow alone. It is important to realise that these devices %%@
operate in a regime where at best the magnetic Reynolds number is close to the marginal one for the laminar flow; it is %%@
definitely below the critical $Rm$ for the turbulent part.  In all the cases discussed the role of turbulence is to %%@
hinder the dynamo, and this can be understood in terms a renormalised diffusivity that increases with the degree of %%@
randomness. This increase in diffusivity in astrophysical situations is irrelevant, since $Rm$ is so far above critical %%@
that increasing the diffusivity makes little difference. However in the case of the experiments, this increase has a %%@
catastrophic effect on the chances of success for the experiment\footnote{ Therefore we conclude that all funding for %%@
experiments should immediately be channeled to theorists studying dynamos at large $Rm$ (or writing reviews)}.

\subsection{Saturation of turbulent dynamos}

The exponential growth of the magnetic field described in the last section is accompanied by an
exponential growth of the magnetic forces, which will eventually become comparable with
those driving the turbulence. 
In this second nonlinear phase the exponential growth of
the magnetic field will become saturated and the magneto-turbulence will settle down to
some stationary level of magnetization. It is of interest to speculate about the nature of the saturation process, and %%@
about the 
general properties of the final state, although in general the saturation mechanism may be subtle (see e.g. \cite{catttob:2009})

As before there is a large difference between high and low magnetic Prandtl number regimes. In the high $Pm$ regime the %%@
dynamo is operating at scales in the sub-inertial range of the velocity for which the Reynolds number is small. %%@
Therefore the inertial term in the
momentum equation can be neglected and the velocity can be split into two components; one driven by the external forces, %%@
which is the original velocity of the kinematic dynamo problem and has a characteric scale $\gg l_{\eta}$, the other, %%@
driven by the Lorentz force, encodes the back-reaction of the magnetic field and is mostly concentrated around the %%@
diffusive scale.
In this high $Pm$ regime saturation can be successfully addressed by semi-analytical models, phenomenological models and %%@
numerical experiments.  The semi-analytical models
are ultimately based on some closure of the MHD equations.
 Within the framework of the Kazantsev equation the magnetically driven velocity produces a change in the velocity %%@
correlator, which leads to the nonlinear saturation (\cite{sub:1999,sub:2003}). A similar approach can be taken by %%@
constructing a Fokker-Planck equation for  the probability distribution function for magnetic fluctuations rather than %%@
an equation for the magnetic correlator. It can be shown that the coefficients of this equation again are determined by %%@
the velocity correlation function which can be modified nonlinearly in a similar manner to above (\cite{bol:2001}). 

An interesting phenomenological model has been proposed by Schekochihin and collaborators (\cite{scheketal:2002-amodel}). %%@
The model begins with the kinematic growth of fields at the resistive scale, which is much smaller than the viscous %%@
scale. In this kinematic phase the eddies driving the dynamo are the ones at the viscous scale. The authors anticipate %%@
that nonlinear effects begin to be important when $\bv \cdot \nabla \bv \approx \bB \cdot \nabla \bB$ at that scale. The %%@
left hand side is easily estimated to be $v^2/l_\nu$. To estimate the right hand side the authors use the foliated %%@
structure of the magnetic field.
To have a geometrical understanding of what this means, it is useful to distinguish between the orientation and the %%@
direction of a vector field. For example  a change from horizontal to vertical is a change in orientation, whereas a %%@
change from up to down is a change in direction. For a typical foliated field the orientation changes slowly on the %%@
scale of the velocity, whereas the direction changes rapidly on the scale $l_\eta$ (\cite{finnott:1988}). Hence the %%@
regions of high curvature occupy practically none of the volume and the right hand side is estimated to be $b^2/l_\nu$, %%@
where $b$ is a typical field strength at the scale 
$l_\eta$. Thus the nonlinear saturation begins when the magnetic energy comes into equipartition with the energy at the %%@
viscous scale. The effect is to suppress the dynamo growth associated with the eddies at the viscous scale. This %%@
suppression is not necessarily achieved by a dramatic reduction of the amplitude of the eddies but rather by a subtle %%@
modification of their geometry. In particular if the velocity becomes more aligned with the local magnetic field then it %%@
cannot distort that field and therefore contribute to its amplification. However, slightly larger eddies can still %%@
sustain growth, albeit at a slightly slower rate. Growth will continue until the magnetic energy comes into %%@
equipartition with the energy of these eddies. The process will continue to larger and larger eddies until the magnetic %%@
energy reaches equipartition with the energy of the flow at the integral scale. This nonlinear adjustment is %%@
characterised by a growth of the magnetic energy on an algebraic (rather than exponential) time and there is a %%@
corresponding shift of the characteristic scale of the magnetic field to larger scales. This idea of scale-by-scale %%@
modification of the velocity to reach some form of global equipartition can be formalised in terms of either %%@
Fokker-Planck equations (\cite{scheketal:2002-amodel}) or a Kazantsev model which is appropriately modified to take %%@
account of the growing degree of anisotropy and finite correlation time (\cite{scheketal:2004-satu}). There are even %%@
models constructed where the final state does not reach global equipartition but only a fraction of equipartition. %%@
According to Schekochihin {\it et al.} (2002), this occurs when $Pm$ is large but $Pm \le Re^{1/2}$ with  $B^2/U^2 \approx %%@
Pm/Re^{1/2}$.
In this scenario, however, by the time the final state is reached the characteristic scale of the magnetic field is %%@
still smaller than the viscous scale. Schekochihin {\it et al.} (2002) estimate this to be the case when $Pm \gg Re^{1/2}$. 

There have been a number of simulations at moderate to high $Pm$, and within the normal restrictions of numerical %%@
simulations the results seem to conform to this general picture (\cite{maronetal:2004}). There is good evidence that in %%@
the kinematic phase the magnetic spectrum is compatible with the $k^{3/2}$ prediction of the Kazantsev model. The %%@
appearance of the magnetic field is indeed that of a foliated structure. In the saturated state, the magnetic and %%@
kinetic energies are comparable, although for moderate values of $Pm$ the magnetic energy increases with $Pm$. As the %%@
saturation progresses the magnetic spectrum grows and flattens, which is compatible with the creation of magnetic %%@
structures larger than the resistive scale.
It is always the case that the magnetic energy exceeds the kinetic energy at small scales (see e.g. %%@
\cite{caratietal:2006}). Moreover the magnetic field is more intermittent than the velocity --- indeed the pdf for the %%@
velocity field remains close to that of a Gaussian whilst that for the magnetic field is better described by an %%@
exponential --- although the degree of intermittency is reduced in the saturated state as compared with the kinematic %%@
state (\cite{cattaneo:1999}). 

\begin{figure}
\centering
\includegraphics[width=1.0\textwidth]{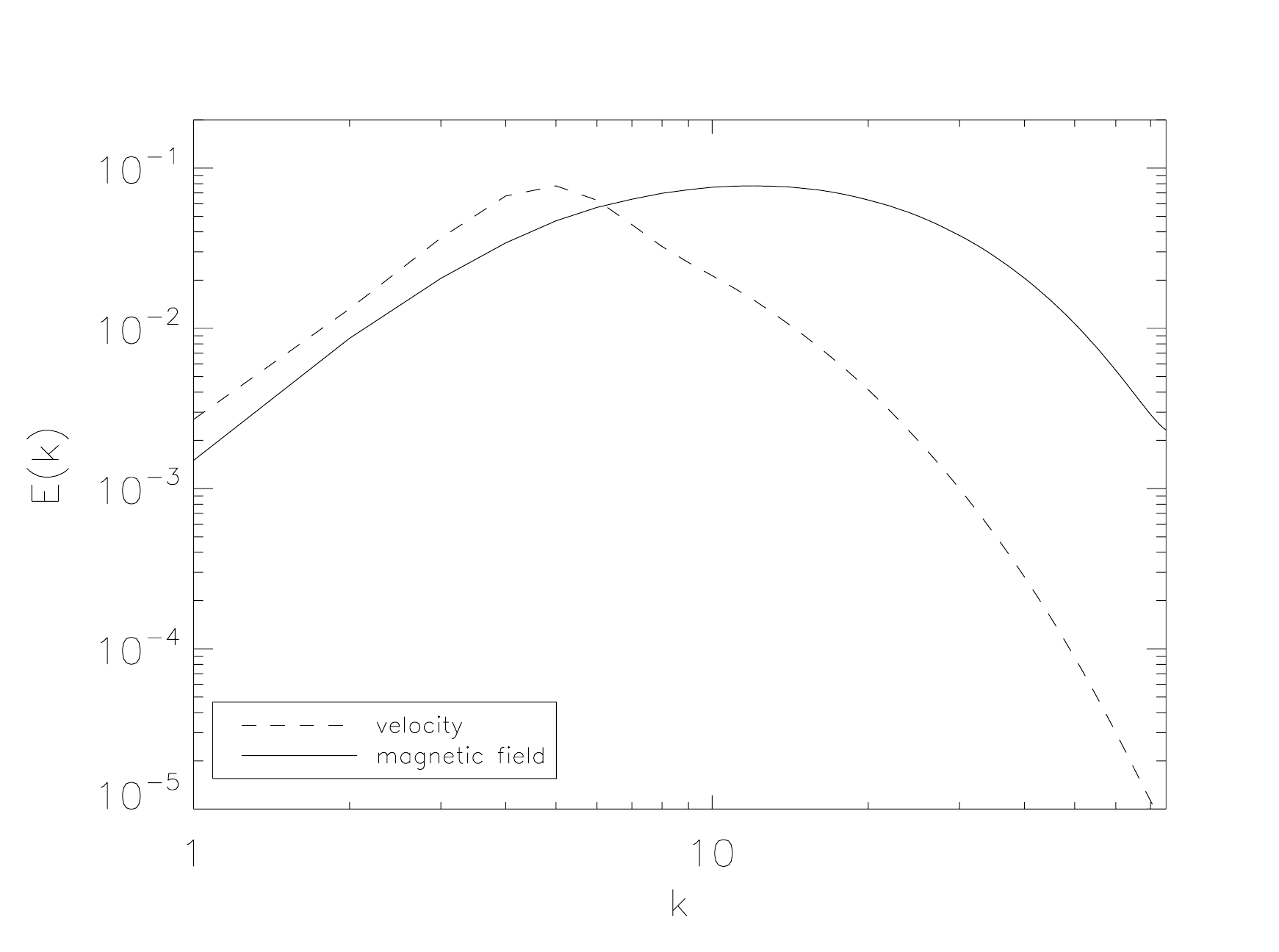}
\caption{Velocity and magnetic spectra in a numerical simulation of magnetised turbulence; courtesy of Mason and Malyshkin (private communication)}
\label{f5}
\end{figure}

\begin{figure}
\centering
\includegraphics[width=1.0\textwidth]{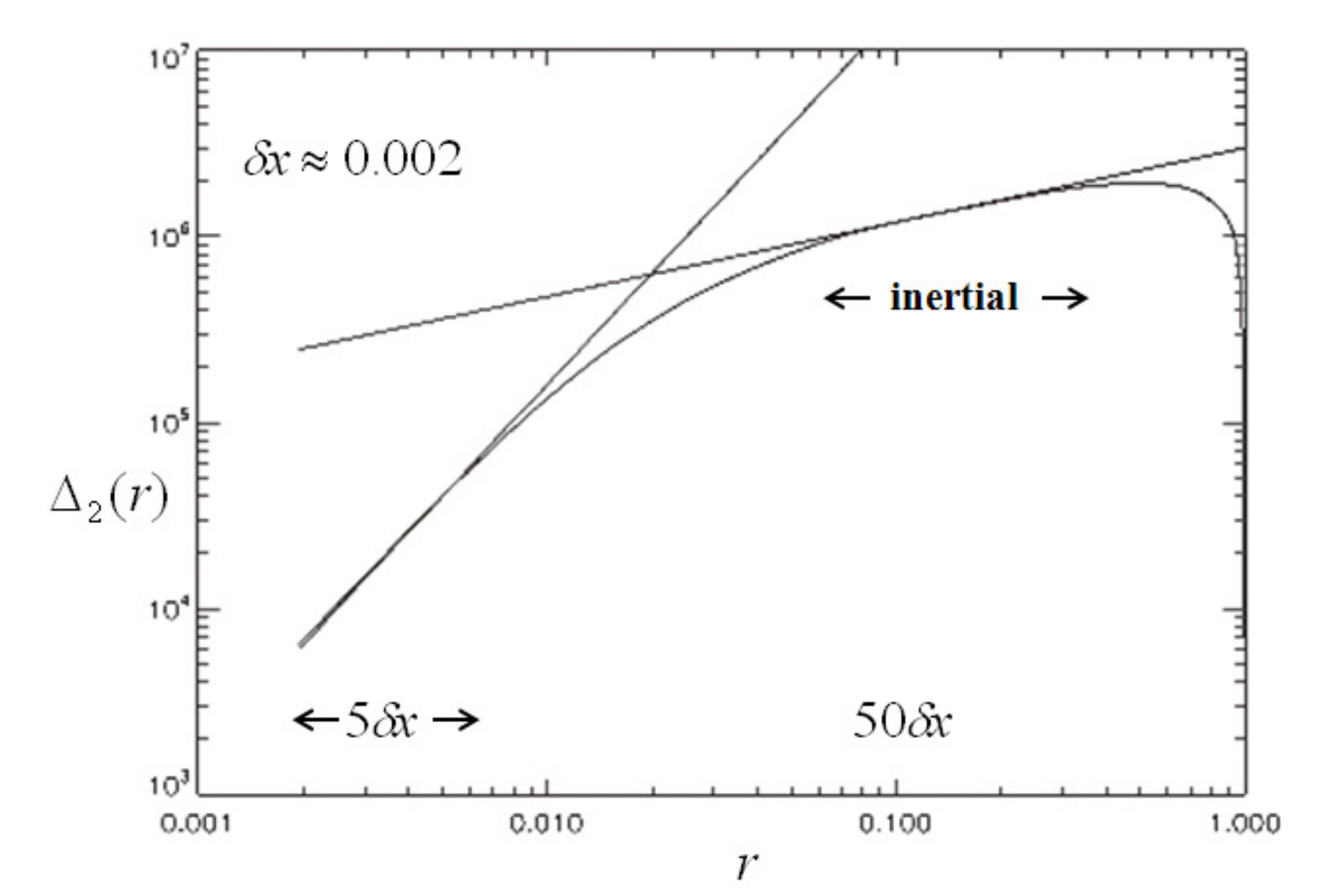}
\caption{Second order longitudinal velocity structure function $\Delta_2(r)$ measured in a numerical simulation of %%@
magnetised turbulence as a function of separation $r$. At small scales $\Delta_2(r) \sim r^2$; at larger separations its %%@
power law behaviour defines the velocity roughness. Clearly, in order to represent numerically a rough velocity, %%@
at least of the order of a few tens of gridpoints is required. Calculations courtesy of Bodo \& Cattaneo.}
\label{f6}
\end{figure}

Remarkably many of these features of the saturated state persist in the $Pm \approx {\cal O}(1)$ regime, which is much %%@
more accessible to numerical simulations. These address various issues ranging from the degree of intermittency to the %%@
nature of the kinetic and magnetic energy spectrum. Although there is no universal agreement there are a number of %%@
features that are robust. The magnetic fluctuations are always more intermittent than the velocity fluctuations. %%@
Furthermore there is a statistical anisotropy along and transverse to the local magnetic field, with a stronger degree %%@
of intermittency transverse 
(\cite{mullbisgrapp:2003}). These authors have argued that the intermittency results are consistent with a She-Leveque %%@
scaling in which the more singular structures are current sheets. Concerning energy spectra, results show that the %%@
kinetic and magnetic energy spectra track each other, with the magnetic energy spectra exceeding the kinetic one at all %%@
but the largest scales, as in Figure~(\ref{f5}). The existence of an inertial range, characterised by a well-defined %%@
power law, is still a matter for debate. There are various reasons for
this. One, of course, is that the resolution of the %%@
numerical simulations is limited. Another one is the presence of bottleneck effects, whose intensity is linked to the %%@
use of hyperdiffusivities. This, coupled with the limited resolution, can have effects that propagate into the putative %%@
inertial range. Finally there is a tendency to present results sometimes in terms of the shell-averaged spectrum, %%@
sometimes in terms of the one-dimensional spectrum relevant to a direction chosen for computational convenience and %%@
sometimes in terms of a one-dimensional spectrum perpendicular to the local mean field. All these issues %%@
notwithstanding, there are two things that can be said with certainty. Most researchers agree that the spectrum, if it exists, has a slope somewhere %%@
between $-1.5$ and $-1.7$ --- we shall return to this issue in the next section. The second is that people have a %%@
tendency to find in their results confirmation of their expectations, and to attribute deviations to other effects, such %%@
as intermittency or bottlenecks. For example it is not unheard of for people to interpret a spectrum compatible with a %%@
spectral slope of $-3/2$ as a spectrum of $-5/3$ with corrections induced by a bottleneck (\cite{haugenetal:2004}) or %%@
intermittency (\cite{beresnyaklazarian:2006}). 
No doubt as computers get bigger and numerics get better some kind of consensus will emerge.

At present the low $Pm$ regime remains largely unchartered. In this regime the dynamo is operating in the inertial range %%@
of the turbulence. Inertial terms are certainly important and the velocity can no longer be thought of as consisting of %%@
two components. 
 In this case analytical progress can only be made by resorting to standard closure models such as EDQNM %%@
(\cite{pouquetetal:1976}).  One of the basic ingredients of these theories is the decorrelation time of triple moments. %%@
Whereas in hydrodynamic turbulence there is overwhelming evidence that this quantity is of the order of the turnover %%@
time, there is no consensus for of what this  quantity should be in  MHD theory. Indeed it may depend sensitively on the %%@
field strength and magnetic Reynolds number of the magnetised turbulence. For this reason it is not clear that utilising EDQNM is
a reliable way to make analytical progress. One of the problems is that, whereas in %%@
hydrodynamic turbulence there is a vast body of experimental evidence, there is no such thing in MHD, and therefore one %%@
is forced to rely almost entirely on numerical experiments, with all of their limitations. These limitations become %%@
extremely severe in the low $Pm$ regime. There is a generic expectation that for sufficiently low $Pm$ an asymptotic %%@
regime is reached where the level of saturation becomes independent of $Pm$ (\cite{petfauve:2008}). Some steps have been %%@
taken to move in the direction of simulating dynamos at low $Pm$, but these are still in the $Pm \sim {\cal O}(1)$  regime, %%@
with all the associated dynamics (see e.g. \cite{mininni:2006,iskakovetal:2007}). It is important to appreciate the %%@
difficulties inherent in reaching the low $Pm$ regime numerically, and why caution should be exercised when interpreting %%@
the results of numerical simulations that are described as being in this regime. The essence of the problem is that the %%@
velocity field driving the dynamo is rough. It is possible to ask what sort of resolution is required to represent such %%@
a velocity. Figure~(\ref{f6}) shows the second order structure function for a turbulent velocity; the distance here is %%@
expressed in units of the grid spacing. At small distances the slope approaches two, implying that on a few %%@
grid-points the function becomes {\it smooth}. It is important to realise that this feature persists even if the %%@
dissipation is entirely numerical. For larger distances (i.e.\ more gridpoints) the slope becomes less than two and %%@
eventually approaches a power law behaviour, characteristic of a rough velocity. So for example, in this case the range %%@
of scales over which the function becomes rough is of the order of 20-30 gridpoints. In the low $Pm$ regime we should %%@
think of this size as the thickness of a magnetic boundary layer that is able to recognise the velocity as rough. %%@
Inspection of Figure~(\ref{fig_2}b) gives an estimate of the size of the domain required to contain a growing eigenfunction %%@
in units of the magnetic boundary layer thickness. For a Kolmogorov-like velocity $1+\alpha = 1.333$ and so this is of the %%@
order of 30. This would give of the order 600-900 gridpoints simply to capture the magnetically growing eigenfunction. %%@
If one then wants a reasonable description of the velocity inertial range one probably requires at least a decade below %%@
this, giving a requirement in terms of gridpoints of several thousands (or tens of thousands). Possibly one could save a %%@
factor of ten by not matching the velocity inertial range to a resolved viscous sub-range by using LES. This is indeed %%@
the approach used by Ponty {\it et al.} (2008), but again we caution the reader that playing this game requires exceptional control of %%@
the LES, and even with the LES, simulations of the order of $1000^3$ are unavoidable.
Again, as machines get larger, the task of exploring this regime may not appear so daunting.

%\subsection{From dynamos to MHD turbulence}
%
%
%Hard to separate the problem. Even the dynamo problem in the deep inertial range looks like the guiding field problem!

%% file: meanfield.tex
\section{Mean field} 
We now turn our attention to the case in which there is an externally imposed magnetic field. As mentioned earlier, here the motivations are that in many astrophysically significant %%@
situations the regions of interest are threaded by a large scale magnetic field, or  that we are interested in small scales whose dynamics is influenced by a magnetic field at %%@
larger scales. It is reasonable, and customary, to concentrate on the idealized case in which the large scale field is uniform. 
%If the large scale field is strong, in a sense that will be defined presently, there is an equivalence between velocity and magnetic fluctuations. 
We begin by discussing small perturbations to a uniform field; the magnetohydrodynamic equations support linear waves that propagate on the stationary and uniformly magnetized background. We %%@
denote the uniform background magnetic field by ${\bf B}_0$ and choose a coordinate frame with the $z$-axis along~${\bf B}_0$. The magnetic field is then given by ${\bf B}({\bf x},t)={\bf %%@
B}_0+{\bf b}({\bf x}, t)$, where ${\bf b}({\bf x}, t)$ is the fluctuating part. Instead of the guide field ${\bf B}_0$ we will often use the Alfv\'en velocity ${\bf v}_A={\bf B}_0/\sqrt{4\pi %%@
\rho_0}$, where $\rho_0$ is a uniform fluid density. We seek wave-like solutions of the MHD equations for the velocity and magnetic fluctuations in the form ${\bf v}({\bf x}, t)={\bf %%@
v}_k\exp(-i\omega t+i{\bf k}\cdot{\bf x})$ and ${\bf b}({\bf x}, t)={\bf b}_k\exp(-i\omega t+i{\bf k}\cdot{\bf x})$. Recall that we are considering the incompressible case, ${\bf k}\cdot %%@
{\bf v}_k$=0, which rules out acoustic modes. 

The remaining waves can be classified according to their polarizations.  
There are two  so-called ``shear-Alfv\'en'' waves.  Their dispersion relation is~$\omega=|k_z|v_A$, and their polarization vector~${\bf b}_k$ is normal to both the wavevector~${\bf k}$ and %%@
the guide field~${\bf B}_0$ (i.e.\ ${\bf b}_k \cdot {\bf k} = {\bf b}_k \cdot {\bf B}_0 = 0$). If the group velocity of a shear-Alfv\'en wave is in the direction of the guide field, its %%@
velocity polarisation is anti-parallel to the magnetic polarisation, i.e. ${\bf v}_k =- {\bf b}_k$. If the group velocity of such a wave is in the direction opposite to the guide field, its %%@
velocity polarisation is parallel to the magnetic polarisation, i.e. ${\bf v}_k= {\bf b}_k$.  

The other two modes are the so-called ``pseudo-Alfv\'en'' waves. Their dispersion relation is also~$\omega=|k_z|v_A$, however, the polarisations are different. The magnetic polarisation is %%@
normal to the wavevector (i.e.\ ${\bf b}_k \cdot {\bf k} = 0$) and lies in the plane of~${\bf k}$ and~${\bf B}_0$. If the group velocity of a pseudo-Alfv\'en wave is in the direction of the %%@
guide field, its velocity polarisation is anti-parallel to the magnetic polarisation, i.e., ${\bf v}_k=- {\bf b}_k$. If the group velocity of such a wave is in the direction opposite to the %%@
guide field, its velocity polarisation is parallel to the magnetic polarisation, i.e., ${\bf v}_k= {\bf b}_k$.   

For those familiar with magnetoacoustic waves, it is helpful to note that the shear-Alfv\'en and pseudo-Alfv\'en linear waves are the limiting cases of the so-called ``Alfv\'en'' and %%@
``slow'' waves, respectively, in the general picture of compressible MHD in the limit of low compressibility, that is $v_A\ll v_s$, where $v_s$ is the speed of sound. An arbitrarily small %%@
perturbation of a uniformly magnetised fluid is a superposition of non-interacting MHD waves. In the purely incompressible case, however, both the shear-Alvf\'en and pseudo-Alfv\'en modes %%@
become rather significant. As we show in the next section, shear-Alfv\'en and pseudo-Alfv\'en wave packets even at finite amplitudes become {\em exact} solutions of the incompressible nonlinear MHD %%@
equations. Any perturbation of incompressible magnetized fluid can be expanded in these modes.   

\subsection{Formulation}
The MHD equations take an especially simple form when written in the so-called Elsasser variables, ${\bf z}^\pm={\bf v}\pm {\bf b}$, 
\begin{equation}
  \left( \frac{\partial}{\partial t}\mp{\bf v}_A\cdot\nabla\right) {\bf 
  z}^\pm+\left({\bf z}^\mp\cdot\nabla\right){\bf z}^\pm = -\nabla P+\frac{1}{2}(\nu +\eta)\nabla^2{\bf z}^{\pm}+\frac{1}{2}(\nu -\eta)\nabla^2{\bf z}^{\mp}   +{\bf f}^{\pm},
  \label{mhd-elsasser}
\end{equation}
where ${\bf v}$ is the fluctuating plasma velocity, ${\bf b}$ is
the fluctuating magnetic field normalized by $\sqrt{4 \pi \rho_0}$,
${\bf v}_A={\bf B}_0/\sqrt{4\pi \rho_0}$ is the contribution from the 
uniform magnetic field ${\bf B}_0$, $P=(p/\rho_0+b^2/2)$ includes the
plasma pressure $p$ and the magnetic pressure $b^2/2$, and the forces ${\bf f}^{\pm}$ mimic possible driving mechanisms. In the case of incompressible fluid, the pressure term is not an independent function, but it ensures incompressibility of the ${\bf z}^+$ and ${\bf z}^-$ fields. Turbulence can be excited in a number of ways, for example by driving velocity fluctuation (the ``dynamo-type'' driving), or by launching %%@
colliding Alfv\'en waves, etc. The steady-state inertial interval should be independent of the details of the 
driving.    

In the incompressible case, it follows from
these equations that for ${\bf z}^\mp({\bf x},t) \equiv 0$, neglecting dissipation and driving, there is an exact
nonlinear solution that represents a non-dispersive wave packet propagating
along the direction $\mp {\bf v}_A$, i.e.\  ${\bf z}^\pm({\bf x},t)=F^\pm({\bf x}\pm {\bf v}_A t)$ where $F^\pm$ is an arbitrary function. A wave packet ${\bf z}^\pm$ therefore
propagates without distortion until it reaches a region where ${\bf z}^\mp$ does not vanish. Nonlinear interactions are thus solely the result
of collisions between counter-propagating Alfv\'en wave packets. These exact solutions may therefore be thought of as typical nonlinear structures in incompressible MHD turbulence, somewhat %%@
analogous to eddies in incompressible hydrodynamic turbulence.\footnote{For other types of turbulence, e.g. rotating or stratified, there are no corresponding exact nonlinear solutions, and the interactions do not occur via collisions in this manner. Care must therefore be taken in applying the ideas and techniques of MHD turbulence more generally.} Any perturbations of the velocity and magnetic fields can be rewritten as perturbations of the Elsasser fields. The %%@
conservation of energy and cross-helicity by the ideal incompressible MHD equations discussed earlier is equivalent to the conservation  of the Elsasser energies $E^+=\int ({\bf %%@
z}^+)^2\,d^3x$ and $E^-=\int ({\bf z}^-)^2\,d^3x$. 
The energies $E^+$ and $E^-$ are independently conserved and they cascade in a turbulent state towards small scales owing to nonlinear interactions of oppositely moving $z^+$ and $z^-$ %%@
Alfv\'en packets.  

Incompressible MHD turbulence therefore consists of  Alfv\'en wave packets, which are distorted and split into smaller ones owing to nonlinear interaction, until their energy is converted %%@
into heat by dissipation. However, when the amplitudes of the wave packets are small they can survive many independent interactions before getting destroyed, in which case MHD turbulence %%@
can be considered as an ensemble of weakly interacting linear waves. 

To estimate the strength of the nonlinear interaction, we need to compare the linear terms, $({\bf v}_A\cdot\nabla) {\bf z}^\pm$, which describe advection of Alfv\'en wave packets along the %%@
guide field, with the nonlinear terms, $\left({\bf z}^\mp\cdot\nabla\right){\bf z}^\pm$, which are responsible for the distortion of wave packets and for the energy redistribution over %%@
scales. Denote $b_{\lambda}$ as the rms magnetic fluctuation at the field-perpendicular scale $\lambda \propto 1/k_{\perp}$, and assume that the typical field-parallel wavenumber of such %%@
fluctuations is~$k_\|$. For Alfv\'en waves, magnetic and velocity fluctuations are of the same order, $v_\lambda \sim b_\lambda$, so one estimates $({\bf v}_A\cdot\nabla) {\bf z}^\pm \sim %%@
v_A k_\| b_\lambda$ and $\left({\bf z}^\mp\cdot\nabla\right){\bf z}^\pm \sim k_\perp b_\lambda^2$.  Then the turbulence is called ``weak'' when the linear terms dominate, i.e.\
\begin{eqnarray}
k_\| v_A\gg k_{\perp} b_{\lambda},
\label{weak-turb}
\end{eqnarray}  
and it is called ``strong''  otherwise. Depending on the driving force, turbulence exhibits either a weak or strong regime in a certain range of scales. In the next section we describe early %%@
approaches to MHD turbulence that essentially treated the turbulence as weak.

%The equations in the introduction support linear waves of the form blah.
%These are the same beast as inertial waves and gravity waves. Small amplitude.
%Shear and Pseudo.

%However there is another property of the incomp MHD equations. In order to see this 
%formulate in terms of Elsasser variables.

%Notation

%Reduced MHD

%Get fully nonlinear solutions that propagate without change of form.

%Given this these solutions the only nonlinear interaction is  with solutions coming in the other direction.

%This leads to the description of MHD turbulence in terms of Alfven packets. NOT EDDIES. eddies splitting becomes wave packets splitting etc.

% Characterise weak and strong by interactions. Linear term

\subsection{Early Approaches}
The first model of MHD turbulence was proposed by Iroshnikov (1963)
and Kraichnan (1965) who invoked the picture of interacting Alfv\'en packets propagating along a strong %%@
large-scale magnetic field.  
%They deduced that the energy transfer time from wave packets  
%of size~$\lambda$ to smaller ones is increased  compared to the simple 
%dimensional Kolmogorov-like estimate $\tau(\lambda)\sim \lambda/v_{\lambda}$.  
Iroshnikov and Kraichnan considered two wave packets of size~$\lambda$ 
propagating with the Alfv\'en velocity in opposite directions along a magnetic-field line.  
They further assumed that the wave packets are isotropic; that is, their field-parallel scale is also~$\lambda$. Assuming that significant interaction occurs between eddies of comparable sizes\footnote{This is a standard assumption of ``locality'' of turbulence, see, e.g., (\cite{aluie_eyink10}).},   
one then estimates from equations~(\ref{mhd-elsasser}) that during one collision, that is one crossing time $\lambda/v_A$, the 
distortion of each wave packet is $\delta v_{\lambda}\sim ( v_{\lambda}^2/\lambda)(\lambda/v_A)$. 
The distortions add up randomly, therefore, each wave packet will be distorted relatively strongly after 
$N\sim (v_{\lambda}/\delta v_{\lambda})^2\sim (v_A/v_{\lambda})^2$ collisions with uncorrelated wave packets moving in the opposite direction. The energy transfer time is, therefore, 
\begin{eqnarray}
\tau_{IK}(\lambda)\sim N\lambda/v_A\sim \lambda v_A/v_{\lambda}^2= \lambda/v_{\lambda}(v_A/v_{\lambda}). 
\label{tauik}
\end{eqnarray} 

It is important that in the Iroshnikov-Kraichnan interpretation,   
a wave packet has to experience {\em many} uncorrelated 
interactions with oppositely moving wave packets (since $\tau_{IK}\gg \lambda/v_A$) before its 
energy is transferred to a smaller scale. The requirement of constant energy flux over 
scales $\epsilon = v^2_{\lambda}/\tau_{IK}(\lambda)={\sf const}$ immediately 
leads to the scaling of the fluctuating fields $v_{\lambda}\propto b_{\lambda}\propto  
\lambda^{1/4}$, which results in the energy spectrum (the  
Iroshnikov-Kraichnan spectrum), 
\begin{eqnarray}  
E_{IK}(k)\sim |v_k|^2k^2\propto k^{-3/2}.
\label{eik}
\end{eqnarray} 
Substituting the derived scaling of the fluctuating fields into formula~(\ref{weak-turb}) one concludes that, since $k_\|\sim k_\perp \sim 1/\lambda$,  turbulence 
becomes progressively {\em weaker} as the scale of the fluctuations decreases.  Therefore, the Iroshnikov-Kraichnan picture is the picture of weak MHD turbulence.

Iroshnikov and Kraichnan did not consider anisotropy of the 
spectrum, so~$E_{IK}(k)$ was assumed to be 
three-dimensional and isotropic. As discussed earlier, over the years the assumption of isotropy was shown to be incorrect. The phenomenological picture of anisotropic %%@
weak MHD turbulence was proposed by Ng \& Bhattacharjee (1996) and
Goldreich \& Sridhar (1997), and it was put on more formal
mathematical ground by Galtier {\it et al.} (2000). This theory is %%@
discussed in the next section. \\
%I-K theory. Cascade time.
%Does not take into account anisotropy

\subsection{Weak turbulence}
\label{weak_turbulence}
When the   
condition~(\ref{weak-turb}) is satisfied, one may assume that
turbulence consists of weakly interacting shear-Alfv\'en and pseudo-Alfv\'en  
waves. The main results of the weak turbulence theory can be explained by using the following dimensional arguments, which were proposed by Ng \& Bhattacharjee (1996) and
Goldreich \& Sridhar (1997) before the more rigorous treatment by  Galtier {\it et al.} (2000) was developed. First, let us note that two Alfv\'en waves with wavevectors ${\bf k}_1$ and ${\bf k}_2$ interact with %%@
a third one if the following resonance conditions are satisfied: $\omega(k)=\omega^+(k_1)+\omega^-(k_2)$ and ${\bf k}={\bf k}_1+{\bf k}_2$. Only the waves propagating in opposite directions %%@
along the guide field interact with each other, therefore, $k_{1\|}$ and $k_{2\|}$ should have opposite signs. Since the Alfv\'en waves have a linear dispersion relation, $\omega=|k_\| |v_A$, %%@
the solution of the resonance equations is nontrivial only if either $k_{1\|}=0$ or $k_{2\|}=0$ (\cite{shebalin83}). The turbulence dynamics therefore depend on the fluctuation spectrum at $k_\|=0$. As the %%@
Alfv\'en waves interact with the fluctuations at $k_\|=0$, the~$k_\|$ components of their  wavevectors do not change, and the energy of waves with given $k_\|$ cascade in the direction of %%@
large~$k_\perp$. 

In the region $k_\perp\gg k_\|$, the polarization of the pseudo-Alfv\'en waves is almost parallel to the guide field ${\bf B}_0$. Therefore, the pseudo-Alfv\'en modes (${\bf z}_p$) influence %%@
the shear-Alfv\'en ones (${\bf z}_s$) through the term $({\bf z}^\pm_p\cdot  \nabla){\bf z}^{\mp}_s \sim z^\pm_p k_\| z^{\mp}_s$, whilst the shear-Alfv\'en modes, whose polarization is %%@
normal to the guide field, interact with each other as $({\bf z}^\pm_{s}\cdot\nabla){\bf z}^{\mp}_s\sim z^{\pm}_sk_\perp z^{\mp}_s$. Since $k_\|$ does not change in the energy cascade, in %%@
the inertial interval we have $k_\perp \gg k_\|$, and we obtain the important property of MHD turbulence that the dynamics of the shear-Alfv\'en modes decouple from the dynamics of the %%@
pseudo-Alfv\'en ones (\cite{goldreich_s95}). Pseudo-Alfv\'en modes are passively advected by the shear-Alfv\'en ones. The spectrum of a passive scalar is the same as the %%@
spectrum of the advecting velocity field (\cite{batchelor:1959}), therefore the spectra of pseudo- and shear-Alfv\'en waves should be the same.

These spectra can be derived from dimensional arguments similar to those of Iroshnikov and Kraichnan, if one takes into account the anisotropy of wave packets with respect to the guide %%@
field. As before, denote the field-perpendicular scale of the interacting wave packets as~$\lambda$, and their field-parallel scale as~$l$; but now the  field-parallel scale~$l$ does not %%@
change during interactions. The collision time (crossing time) is now given by~$l/v_A$. During one collision,  the  counter-propagating packets get deformed by $\delta v_\lambda \sim %%@
(v_\lambda^2/\lambda)(l/v_A)$. The number of uncorrelated interactions before the wave packets get destroyed is $N\sim (v_\lambda/\delta v_\lambda)^2\sim \lambda^2 v_A^2/(l^2 v_\lambda^2)$, %%@
and the energy cascade time is $\tau_w\sim Nl/v_A$. From the condition of constant energy flux $\epsilon=v_\lambda^2/\tau_w={\rm const}$, one derives $v_{\lambda}\propto b_\lambda\propto %%@
\lambda^{1/2}$, and the field-perpendicular energy spectrum 
\begin{eqnarray}
E(k_\perp)\propto k_\perp^{-2},
\label{weak-spectrum}
\end{eqnarray}
where the two-dimensional Fourier transform is used in the $(x,y)$ plane in order to calculate the spectrum.

These results can be put on a more rigorous mathematical ground by using the approach of weak turbulence theory. The basic assumption of this theory is that 
in the absence of nonlinear interaction, the waves have random phases, and that 
the Gaussian rule can be applied to express their higher order  
correlation functions in terms of the second-order ones.\footnote{Many 
  papers contributed over the years to the development of fundamental
  ideas on MHD turbulence (see e.g., the reviews 
  in~\cite{biskamp:2003,ng03}). The general methods of weak turbulence theory
  have been extensively reviewed (\cite{zakharov,newell}).} 
A perturbative theory of weak MHD turbulence has also been developed (\cite{galtier_nnp00}, see also \cite{galtier_nnp02}). 
By expanding the Elsasser form of the MHD eq.~(\ref{mhd-elsasser}) up 
to the second order in the nonlinear interaction and using the
Gaussian rule to split the fourth-order correlators, the authors  
derived a closed system of kinetic equations governing 
the wave energy spectra\footnote{It should be noted that the Gaussian assumption is, in fact, not necessary to derive the kinetic equatons. The same closure for the long time behavior of
the spectral cumulants can be derived without a priori assumptions on the statistic of the process, e.g., Newel {\em et~al.} (2001),  Elskens \& Escande
  (2003).}. 

These equations confirm the important fact 
that wave energy cascades in the Fourier 
space in the direction of large~$k_{\perp}$, and the universal spectrum 
of wave turbulence is established in the region~$k_{\perp}\gg k_{\|}$. The equations also demonstrate that  
in this region the dynamics of the shear-Alfv\'en waves decouple 
from the dynamics of the pseudo-Alfv\'en 
waves, and the pseudo-Alfv\'en waves are passively advected by the
shear-Alfv\'en ones, in agreement with the previous qualitative 
consideration. The main objects of interest in this theory are the correlation functions of the shear-Alfv\'en waves, $\langle {\bf z}^{\pm}({\bf k}, t){\bf z}^\pm({\bf k}', t) \rangle %%@
=e^{\pm}({\bf k}, t)\delta({\bf k}+{\bf k}')$. In addition, it was assumed that the waves propagating in the opposite directions are not correlated, that is, %%@
$\langle {\bf z}^{\pm}({\bf k}, t){\bf z}^\mp({\bf k}', t) \rangle =0$. The kinetic equations for the energy spectra $e^{\pm}({\bf k}, t)$ then have the form: 
\begin{eqnarray}
\partial_t e^{\pm}({\bf k})= \int M_{{\bf k}, {\bf p} {\bf q}}
e^{\mp}({\bf q}) [e^{\pm}({\bf p})-e^{\pm}({\bf k})]\delta(q_\|)
d_{\bf k,pq},
\label{galtier-eq}
\end{eqnarray}
where the interaction 
kernel is given by
\begin{eqnarray}
M_{{\bf k}, {\bf p}{\bf q}}=({\pi}/{v_A}){({\bf
    k}_{\perp}\times{\bf q}_{\perp})^2({\bf k}_{\perp}\cdot {\bf
    p}_{\perp})^2}/({k_{\perp}^2 q_{\perp}^2 p_{\perp}^2}),
\end{eqnarray}  
and the shorthand notation $d_{\bf k,pq}  
\equiv \delta({\bf k}-{\bf p}-{\bf q})\,d^3p\,d^3q $ is adopted. It is also customary to  
use the phase-volume  compensated energy spectrum calculated as $E^{\pm}({\bf k}, t)dk_{\|}\,dk_{\perp}= e^{\pm}({\bf k},
t)2\pi k_{\perp}dk_{\|}\,dk_{\perp}$.  In this section we consider only statistically balanced turbulence, that is, we assume $E^+=E^-$. The balanced stationary 
solution of equation~(\ref{galtier-eq}) was found
(\cite{galtier_nnp00}) to have 
the general form~$E^{\pm}({\bf k})=g(k_{\|})k_{\perp}^{-2}$, where
$g(k_{\|})$ is an arbitrary function that is smooth at~$k_{\|}=0$.

It should be noted, however, that the derivation of~(\ref{galtier-eq}) based
on the weak interaction approximation is not rigorous. 
It follows from the wave resonance condition, and as is evident from equation~(\ref{galtier-eq}),   
that only the $q_\|=0$~components of the energy spectrum $e({\bf q})$ are
responsible for the energy transfer. However, if we apply equation~(\ref{galtier-eq}) to
these dynamically important  
components themselves, that is, if we set $k_{\|}=0$ in (\ref{galtier-eq}), 
we observe an inconsistency. Indeed, the perturbative approach implies
that the linear  
frequencies of the waves are much larger than the 
frequency of 
their nonlinear  
interaction. The nonlinear interaction 
in~(\ref{galtier-eq}) remains nonzero as $k_{\|}\to 0$ while the
linear frequency of  
the corresponding Alfv\'en modes, $\omega_k=k_{\|}v_A$, vanishes. Therefore, as shown 
by Galtier {\it et al.} (2000), an additional assumption that goes beyond the theory of weak turbulence, should be made. Namely, one has to assume the smoothness of the 
function $g(k_\|)$ at $k_\| =  0$; this assumption is crucial for deriving  
the spectrum $E({\bf k}) \propto k_{\perp}^{-2}$ since, according to (\ref{galtier-eq}) the wave dynamics essentially depend on the energy spectrum at~$k_\|\to 0$. 

A definitive numerical verification of such a spectrum  
therefore seems desirable.  It is however quite difficult to perform direct numerical simulations of weak MHD turbulence based on equation~(\ref{mhd-elsasser}). 
The major problem 
faced by such simulations  
is to satisfy simultaneously the two conditions,  
$k_\perp\gg k_{\|}$, which is necessary to reach the universal regime where the dynamically unimportant pseudo-Alfv\'en mode decouples, and $k_{\|}B_0\gg k_\perp b_{\lambda}$, which is the %%@
condition of weak turbulence. These two conditions are hard to achieve with present-day computing power.
\begin{figure} 
\centering
\includegraphics[width=1.0\textwidth]{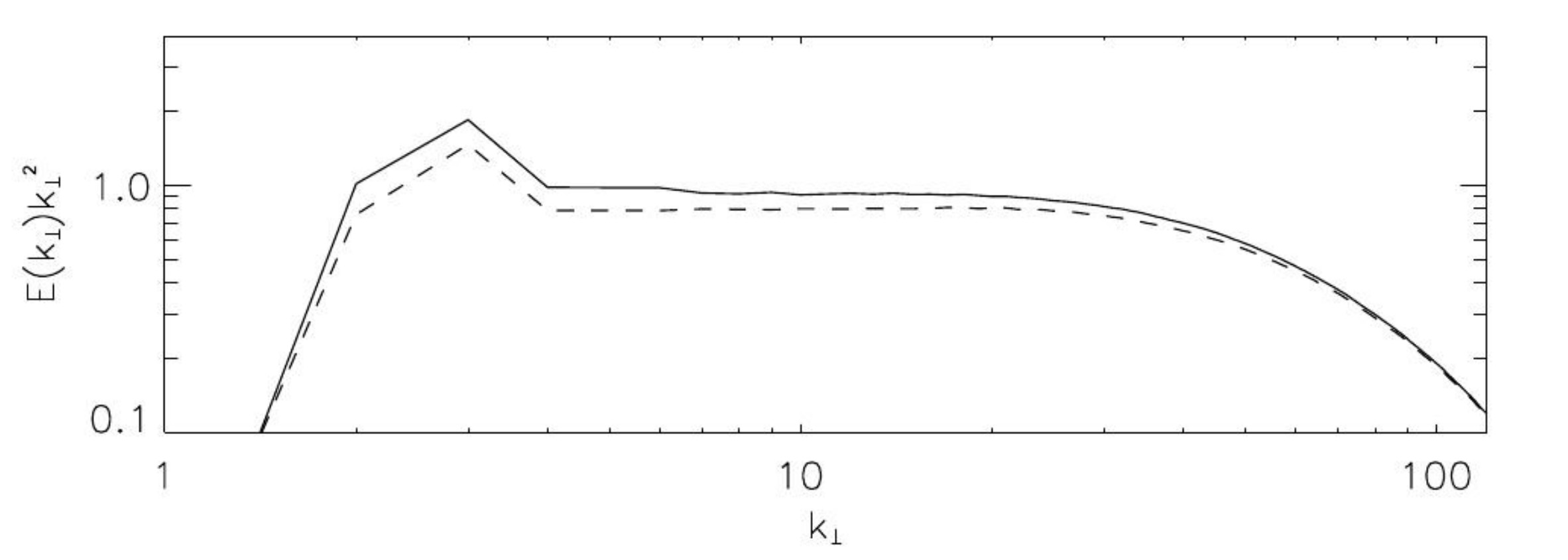}
\caption{Compensated spectra of $E^+$ (solid line) and $E^-$ (dash
  line) in numerical simulations of weak MHD turbulence based on
  reduced MHD equations, with numerical resolution $1024^2\times 256$,
  and Reynolds number ${Re=Rm}=6000$  (from \cite{boldyrev_p09}).}  
\label{balanced_weak}
\end{figure}

The first numerical tests (\cite{bhattacharjee_n01} and \cite{ng03}), used a scattering 
model based on MHD equations expanded up to the second order in 
nonlinear interaction rather than the full MHD equations. Integration of such a model did reproduce 
the~`$-2$ exponent'. Recently, direct numerical simulations of the so-called reduced MHD (RMHD) equations (see later), which explicitly neglect pseudo-Alfv\'en fluctuations and present a good %%@
approximation of the full MHD equations in the case of strong guide
field, were conducted (\cite{perez_b08,boldyrev_p09}); more details on numerical simulations are given in Section~(\ref{numerical_frameworks}). They also confirmed the $-2$ exponent, see Figure~(\ref{balanced_weak})

Weak turbulence theory predicts that as $k_\perp$ increases, the turbulence should eventually become strong. Indeed, owing to the obtained scaling of turbulent fluctuations, %%@
$b_{\lambda}\propto \lambda^{1/2}$, the linear terms in equation~(\ref{mhd-elsasser}) scale as $({\bf v}_A\cdot\nabla){\bf z}^{\pm}\sim v_Ab_\lambda/l\propto \lambda^{1/2}$, whilst the nonlinear %%@
ones are independent of $\lambda$ as $({\bf z}^\mp\cdot \nabla){\bf z}^\pm \sim b_\lambda^2/\lambda \propto {\rm const}$. One therefore observes  that the condition of weak turbulence %%@
(\ref{weak-turb}) should break at small enough~$\lambda$, which means that at large $k_\perp$ turbulence becomes strong.  In the next section  we describe strong MHD turbulence.

%qualitative analysis
%Bhatt \& Ng

%quantitative analysis Galtier et al.. Numerical simulations. Lithwick \& Goldreich solved the integor-diff equations numerically and matched with forcing and %dissipation.

%DNS of reduced MHD. special setting of the force.

%energy propagates in direction of large $k_perp$. If go far in inertial range turbulence becomes strong

\subsection{Strong turbulence}
\label{strong_turbulence}
As noted at the end of the previous section, weak MHD turbulence eventually becomes strong as the $k_{\perp}$ increases so that condition (\ref{weak-turb}) is no longer satisfied. If this is the case then one  can argue that the linear and nonlinear terms should be %%@
approximately balanced at all scales; in equation~(\ref{weak-turb}) this
would mean that  ``$k_\| v_A\sim k_{\perp} b_{\lambda}$''. This is the
so-called critical balance condition (\cite{goldreich_s95}). Before discussing the consequences of critical balance, let us give its more precise definition, and discuss its physical meaning.

In contrast with weak  turbulence, in strong turbulence the magnetic field lines are relatively strongly bent by velocity  fluctuations. The wave packets following these lines in opposite %%@
directions are strongly distorted in only one interaction, that is, one crossing time. During such an interaction small wave packets are guided not by the mean field obtained through averaging %%@
over the whole turbulent domain, but rather by a local guide field
whose direction is deviated from the direction of the mean field by
larger wave packets (\cite{cho_v00}). Therefore, it would be incorrect to verify the critical balance condition by using the wavenumber $k_\|$ defined through the Fourier transform in the global $z$-direction, as sometimes proposed in %%@
the literature. Rather, the critical balance condition means that the nonlinear interaction time $\lambda/b_\lambda$ should be on the order of the linear crossing time, which can be represented as %%@
$l/v_A$ with some length~$l$ along the {\em local} guide field, at all scales.  
%It should be noted however that the local direction of a magnetic field cannot be measured precisely since different magnetic field lines deviate within the angle $\sim \lambda/l$ on the 
% scale of the eddy. 
%Rather, for the verification of 
%critical balance condition, the nonlinear interaction time (nonlinear energy transfer time) should be measured 
%as a function of scale, as e.g., is done in Maron \& Goldreich. 

The physical meaning of the critical balance can then be understood
from the following causality principle (\cite{boldyrev05}).  Suppose that owing to nonlinear interaction the wave packets %%@
are deformed on the time scale $\tau_N\sim \lambda/v_{\lambda}$. During this time, the information about the deformation cannot propagate along the guide field further than a distance %%@
$l\sim v_A\tau_N$, and the fluctuations cannot be correlated at a larger distance along the guide field. This coincides with  the statement of the critical balance. We also note that the %%@
condition of critical balance in the GS picture has a useful geometric property. Noting that the individual  magnetic field lines in an eddy of size $\lambda$ are locally deviated by the small %%@
angle $b_\lambda/v_A$, one derives that the Alfv\'en wave packet of length $l$ displaces the magnetic field lines  
in the field perpendicular direction by a distance~$\xi\sim b_{\lambda}l/v_A $. In the GS picture this displacement happens to be equal to the perpendicular wave-packet size~$\lambda$.

As a consequence of critical balance, oppositely moving Alfv\'en wave-packets are significantly deformed during only one interaction (one crossing time). This is in  contrast with weak %%@
turbulence discussed earlier, where a large number of crossing times was required to deform a wave packet. The nonlinear interaction time therefore assumes the Kolmogorov form $\tau_{GS}\sim %%@
\lambda/v_\lambda$, and the energy spectrum attains the Kolmogorov scaling in the field-perpendicular direction (the Goldreich-Sridhar spectrum):
\begin{eqnarray}
E_{GS}(k_\perp)\propto k_\perp^{-5/3},
\label{gs}
\end{eqnarray}
where the spectrum is calculated by using the two-dimensional Fourier transform in the $(x,y)$ plane, in analogy with the anisotropic weak turbulence spectrum~(\ref{weak-spectrum}). As an %%@
important consequence of critical balance, strong MHD turbulence becomes progressively more anisotropic at smaller scales. Indeed, with the GS scaling $v_\lambda\propto \lambda^{1/3}$ and %%@
the critical balance condition it follows that~$l\propto \lambda^{2/3}$, and the ``eddies'' or wave packets get progressively elongated along the local guide field as their scale decreases. 

\begin{figure} 
\centering
\includegraphics[width=1.0\textwidth]{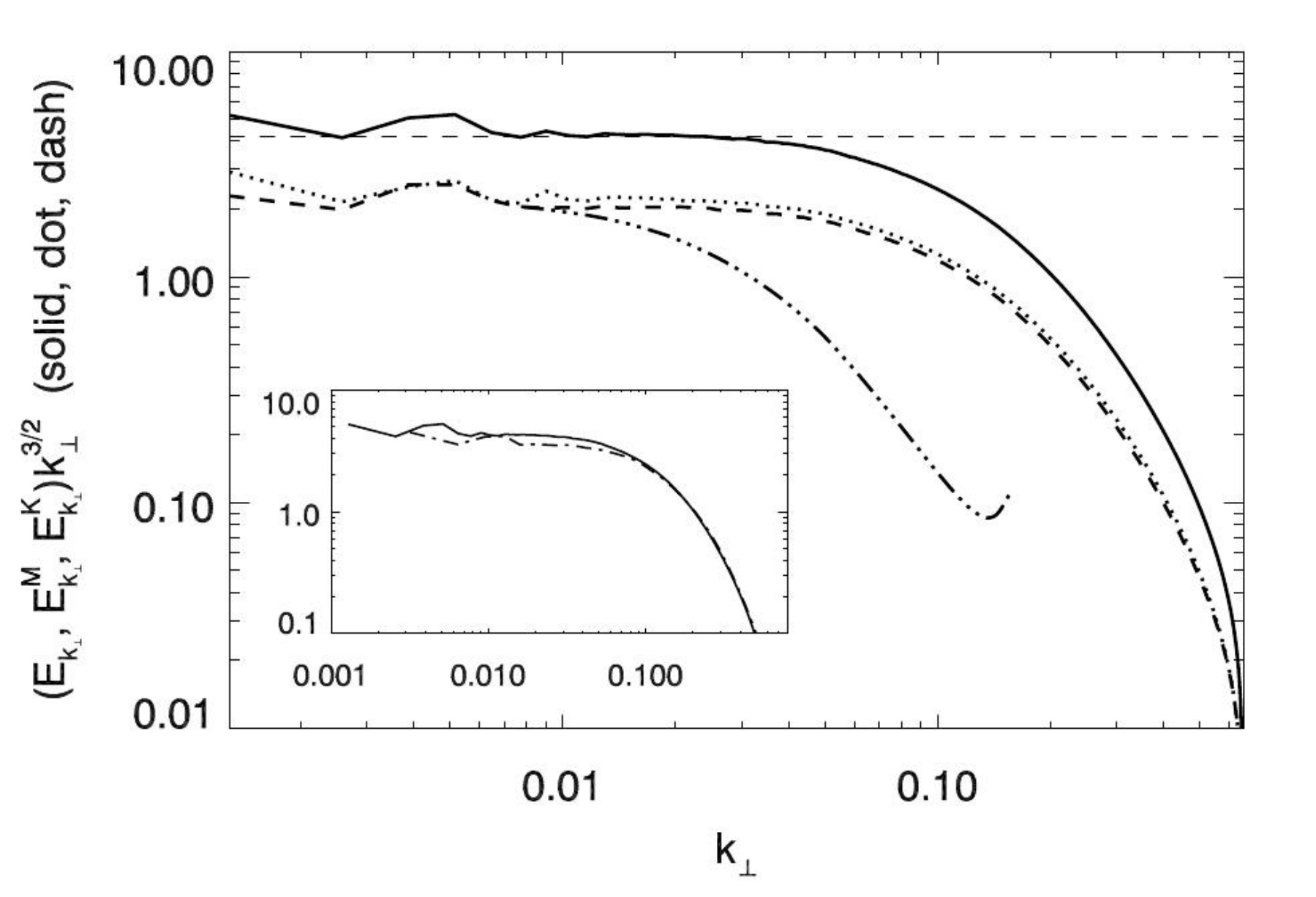}
%\centerline{\psfig{file=muller_grappin2005.eps,width=5.2in,angle=0}}
\caption{Compensated field-perpendicular total (solid), kinetic
  (dashed), and magnetic (dotted) energy spectra. The guide field is
  $B_0=5v_{rms}$, resolution $1024^2\times 256$, Reynolds number
  $Re=Rm=2300$ (based on field-perpendicular fluctuations).
  Dash-dotted curve: high-k part of field-parallel total energy
  spectrum. Insert: comparison of the field-perpendicular energy
  spectra for field-perpendicular resolutions of $512^2$ (dash-dotted)
  and $1024^2$ (solid) (from \cite{muller_g05}).}  
\label{muller_grappin2005}
\end{figure}

Such scale-dependent anisotropy and the scaling (\ref{gs}) seemed to find some support in early numerical simulations, which did not have strong enough guide field, i.e. they had $v_A\sim v_{rms}$, and %%@
did not have large enough inertial interval, (see e.g.,
\cite{cho_v00,cho_lv02}).  Recent high resolution direct numerical simulations with a strong guide field, $v_A\geq 5v_{rms}$, verify the strong anisotropy of the %%@
turbulent fluctuations, supporting the argument that the original IK picture is incorrect. However, they consistently reproduce the field-perpendicular energy spectrum as close to %%@
$E(k_\perp)\propto k_\perp^{-3/2}$, (see, e.g.,
\cite{maron_g01,mullbisgrapp:2003,muller_g05}, Mason {\it et al.} 2006, \cite{mason_cb08,perez_b08}),  thus contradicting the GS model, see Figure (\ref{muller_grappin2005}).

The flattening of the spectrum compared with the GS theory means that the energy transfer time becomes progressively longer than the Goldreich-Sridhar time~$\tau_{GS}(\lambda)$ at small %%@
$\lambda$.  This can happen if the nonlinear interaction in the MHD
equations is depleted by some mechanism. A theory explaining such
depletion of nonlinear interaction has recently been proposed  %%@
(\cite{boldyrev05,boldyrev06}). In addition to the 
elongation of the eddies in the direction of the guiding field, it is proposed that 
%the fluctuating velocity and magnetic 
%fields at a scale $\lambda\sim 1/k_{\perp}$
%are aligned within a small scale-dependent angle in the field perpendicular plane, $\theta \propto \lambda ^{1/4}$. The essence of the phenomenon is that 
at each  
field-perpendicular scale $\lambda$ ($\sim 1/k_{\perp}$) in the inertial range, typical 
shear-Alfv\'en velocity fluctuations, ${\bf v}_{\lambda}$, and magnetic fluctuations, $\pm {\bf b}_{\lambda}$, tend 
to align the directions of their polarizations in the 
field-perpendicular plane, and the turbulent eddies become anisotropic in that plane. In these eddies the  magnetic and velocity fluctuations change significantly in the direction almost %%@
perpendicular to the directions of the fluctuations themselves,~${\bf v}_\lambda$ and~$\pm {\bf b}_\lambda$. This reduces the strength of the nonlinear interaction in the MHD equations by %%@
$\theta_\lambda \ll 1$, which is the angle between the direction of the fluctuations and the direction of the gradient: $({\bf z}^{\mp}\cdot \nabla){\bf z}^{\pm}\sim v_{\lambda}^2 %%@
\theta_\lambda/\lambda$.  
%Numerical illustrations of this phenomenon can be found in \cite{perez_b09,boldyrev_mc09}, see also \cite{matthaeus08}. 
One can argue that the alignment and anisotropy  
are stronger for smaller scales, with the alignment angle decreasing with scale 
as $\theta_{\lambda}\propto \lambda^{1/4}$. This `dynamic alignment'
process progressively reduces the strength of the nonlinear interactions as the scale of the fluctuations decreases, which leads to the velocity and magnetic fluctuations $ v_{\lambda}\sim b_\lambda \propto \lambda^{1/4}$, and to the field-perpendicular energy spectrum  
\begin{eqnarray}
E(k_{\perp}) \propto k_{\perp}^{-3/2}.
\end{eqnarray}

Dynamic alignment is a well-known phenomenon of MHD turbulence, (e.g.,
Dobrowolny {\it et al.} 1980, \cite{biskamp:2003}), but the term has taken on a number of meanings.  
In  early studies it essentially meant that {\em decaying} MHD turbulence asymptotically reaches the so-called Alfv\'enic state where either ${\bf v}({\bf x})\equiv {\bf b}({\bf x})$ 
or ${\bf v}({\bf x})\equiv -{\bf b}({\bf x})$, depending on initial conditions. Such behaviour is a consequence of cross-helicity conservation, (see e.g., \cite{biskamp:2003}):  energy decays faster than cross-helicity, and %%@
such selective decay leads asymptotically to alignment of magnetic and velocity fluctuations. 
Regions of polarized fluctuations have also been previously detected in numerical simulations of {\em driven} turbulence (\cite{maron_g01}). The essense of the phenomenon that we discuss here is that in randomly driven MHD turbulence the fluctuations ${\bf v}_\lambda$ and $\pm  {\bf b}_\lambda$ tend 
to align their directions in such a way that the alignment angle
becomes {\em scale-dependent} (\cite{boldyrev05,boldyrev06,mason_cb06,boldyrev_mc09}).

\begin{figure} 
\centering
\includegraphics[width=1.0\textwidth]{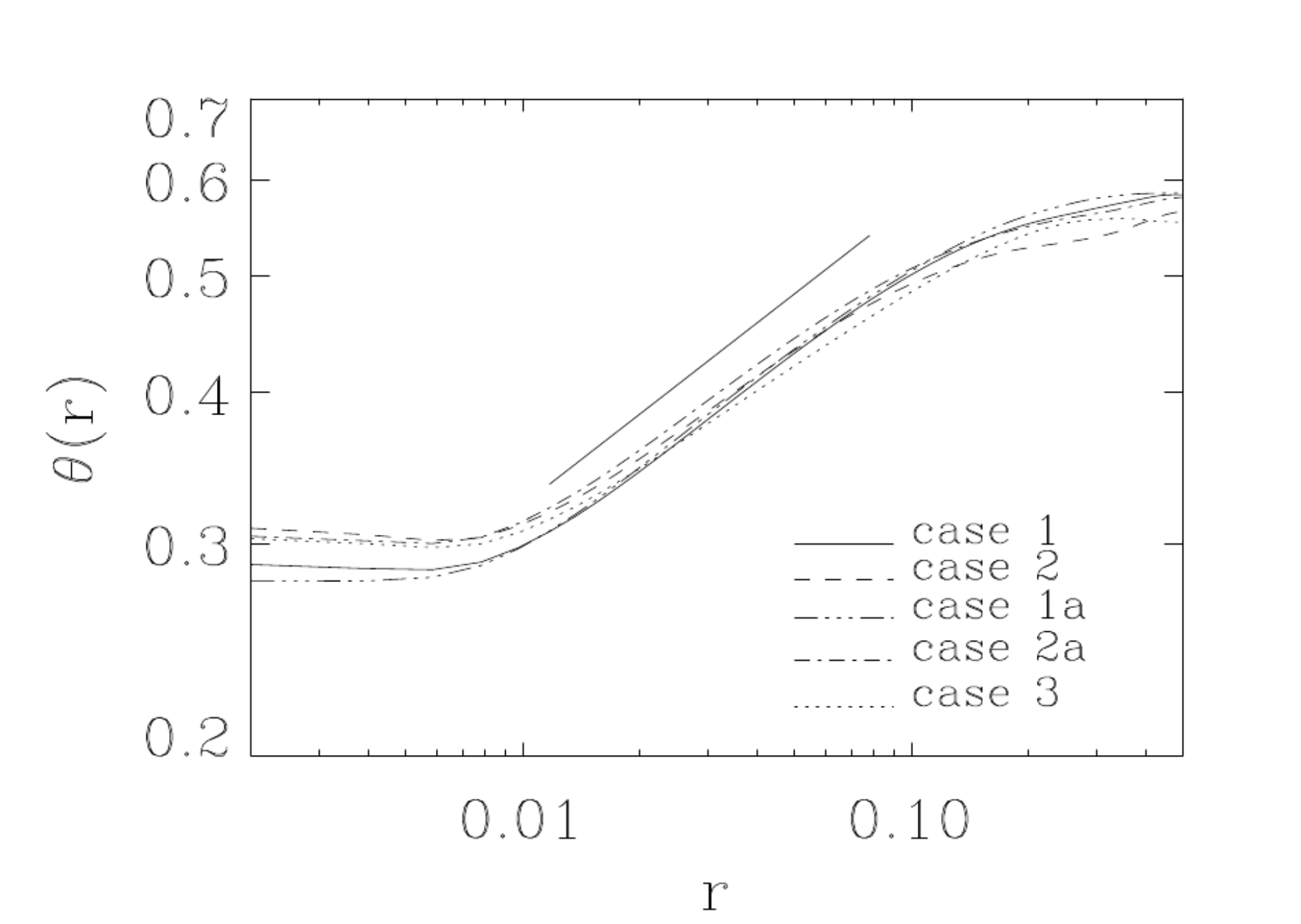}
%\centerline{\psfig{file=alignment_angle.ps,width=5.2in,angle=0}}
\caption{The alignment angle between shear-Alfv\'en velocity and
  magnetic fluctuations as a function of fluctuation scale~$r$. The
  results are plotted for five independent simulations that differ by
  the large-scale driving mechanisms.  The solid line has the slope of
  1/4, predicted by the theory (from \cite{mason_cb08}).}  
\label{alignment_angle}
\end{figure}

The first numerical observations of the scale-dependent dynamic
alignment were presented in Mason {\it et al.} (2006) and Mason {\it et
  al} (2008), (see also Beresnyak \& Lazarian 2006). The results are shown in  Figure
(\ref{alignment_angle}). The effect was also observed in magnetized
turbulence in the solar wind (\cite{podesta09}). It has been also
argued (Boldyrev {\it et al.} 2009) that the phenomenon of scale-dependent dynamic alignment, and the corresponding energy spectrum 
$k^{-3/2}_{\perp}$ are consistent with the exact relations known for
MHD turbulence (\cite{politano_p98}).  

As previously discussed, there are two possibilities for dynamic alignment:  
the velocity fluctuation ${\bf v}_\lambda$ can be aligned either 
with ${\bf b}_\lambda$ (positive alignment) or with $-{\bf b}_\lambda$ (negative alignment). This implies that the turbulent domain can be fragmented into regions (eddies) of positive and negative %%@
alignment. If no overall alignment 
is present, that is, the total cross-helicity of turbulence is zero,
the numbers of positively and negatively aligned eddies  are equal on
average. In the case of nonzero total cross-helicity those numbers may
be not equal. However, there is no
essential difference between overall balanced (zero total
cross-helicity) and imbalanced (non-zero total cross-helicity) strong
turbulence (\cite{perez_b09}). Therein it is argued that strong MHD turbulence, %%@
whether balanced overall or not, has the characteristic property that at each scale it is locally imbalanced. 
%(in the next section we provide a numerical illustration of this %%@phenomenon). 
Overall, it can be viewed as a superposition of positively and negatively aligned eddies.  The scaling of the turbulent energy spectrum depends on the change in degree of alignment with %%@
scale, not on the amount of overall alignment.

%It has been recently demonstrated that the phenomenon of scale-dependent dynamic alignment, and the corresponding energy spectrum %%@
%$k^{-3/2}_{\perp}$ are consistent with the exact relations known for MHD turbulence due to work by \cite{politano_p98}, see \cite{boldyrev_mc09}. 

%Linear term no longer large

%G-S model. Critical balance and anisotropy. If you want critical balance then anisotropy becomes scale-dependent. And crit balance
%determines the degree of anisotropy. Anisotropy of the cascade (not the thing).

%How do you measure anisotropy?

%Early numerical simulations. Lazarian \& Cho, Vishniac observed this anisotropy. Maron \& Goldreich, Grappin, Muller \& Biskamp

%Dynamic alignment: indications that nonlinear interactions were not quite as strong as suggested by G-S. Need some mechanism to weaken the interactions.

%The mystery of the $3/2$-spectrum. Bizarre explanations for it including bottleneck effect. Numerical simulations: Pouquet \& Minnini,  

%Be careful. Need to set numerical simulations up right. Infinite number of ways of getting it wrong.

\subsection{Numerical frameworks}
\label{numerical_frameworks}
In the absence of a rigorous theory, the understanding of MHD turbulence largely relies on phenomenological arguments and numerical simulations. Numerical experiments are sometimes able to %%@
resolve controversies among different phenomenological assumptions.  In this section we discuss the specific requirements for the numerical simulations imposed by the special features of MHD %%@
turbulence. In particular, the anisotropic nature of MHD turbulence requires the optimisation of the computation domain and the driving force in order to capture the dynamics of strongly %%@
anisotropic eddies. First, one needs to choose the forcing as to ensure that large-scale eddies, driven by the forcing, are critically balanced.   Second, the simulation box should be %%@
expanded in the direction of the guide field ($z$-direction) so as to fit the anisotropic eddies, otherwise, residual box size effects can spoil the inertial interval if the Reynolds number 
if not large enough. 

Numerical simulations, which have an inertial interval of a limited extent, demonstrate that when the guide field is not strong enough,  $v_A\leq v_{rms}$, the observed turbulence spectrum %%@
is not very different from the Kolmogorov spectrum (e.g.,
\cite{cho_v00,cho_lv02,mason_cb06}). As~$v_A$ exceeds~$\sim 3
v_{rms}$, the guide field becomes important and the spectral exponent
changes to~$-3/2$, (e.g., \cite{mullbisgrapp:2003}, Mason {\it et al.} 2006).  %%@
In the latter case, the simulation box should have the aspect ratio
$L_\|/L_\perp \sim v_A/v_{rms}$. Current state-of-the-art
high-resolution direct numerical simulations of MHD turbulence
(\cite{muller_g05,perez_b09,perez_b10}) discuss in detail  the aforementioned effects of anisotropy related to the strong guide field. 

We now demonstrate that anisotropic MHD turbulence can, in fact, be effectively simulated using simplified MHD equations. An obvious simplification stems from the fact that in the case of %%@
strong guide field the gradients of fluctuating fields are much smaller along the guide field than across the field. Such a system of equations, the so-called Reduced MHD (RMHD) equations, %%@
was derived in context of fusion devices (\cite{strauss76,kadomtsev_p74}, see also \cite{biskamp:2003}). In this system the following self-consistent ordering is used $L_\perp/L_\| \sim %%@
b_\perp/B_0\sim b_\|/b_\perp \ll 1$, where $L_\|$ and $L_\perp$ are typical scales of magnetic perturbations, i.e., large scales of turbulent fluctuations in our case. As noticed by Goldreich \& Sridhar (1995) this %%@
is precisely the ordering following from the critical balance condition for strong MHD turbulence. Therefore, the Reduced MHD system is suitable for studying strong turbulence. The Reduced %%@
MHD equations have the form:
\begin{eqnarray}   
& \left( \frac{\partial}{\partial t} \mp {\bf v}_A\cdot \nabla \right) {\tilde
    {\bf z}}^{\pm}+({\tilde {\bf z}}^{\mp}\cdot \nabla){\tilde {\bf
      z}}^{\pm}=-\nabla_{\perp} P +\nonumber \\ 
& +\frac{1}{2}(\nu +\eta)\nabla^2{\bf z}^{\pm}+\frac{1}{2}(\nu -\eta)\nabla^2{\bf z}^{\mp} +{\tilde {\bf f}}_\perp ,
\label{rmhd} 
%\nonumber\\ 
% \nabla \cdot {\tilde {\bf z}}^{\pm}&=&0. 
\end{eqnarray}
where the fluctuating fields and the force have only two vector components, 
e.g., ${\tilde {\bf z}}^{\pm}=\{{\tilde z}^{\pm}_1, {\tilde z}^{\pm}_2, 0 \}
$, but depend on all three spatial coordinates. Since both the velocity and the magnetic fields are divergence-free, they can be represented through the scalar potentials, ${\bf v}={\hat %%@
{\bf z}}\times \nabla \phi$ and ${\bf b}={\hat {\bf z}}\times \nabla \psi$, where ${\hat {\bf z}}$ is the unit vector in the z-direction. Then, the system~(\ref{rmhd}) yields two scalar %%@
equations,
\begin{eqnarray}
{}& \partial_t \omega + ({\bf v}\cdot \nabla) \omega -({\bf b}\cdot \nabla)j =B_0\partial_z j +\nu \nabla^2\omega +f_\omega , \nonumber \\
{}& \partial_t \psi +({\bf v}\cdot \nabla) \psi=B_0 \partial_z \phi +\eta \nabla^2 \psi +f_\psi,
\label{rmhd2}
\end{eqnarray}
where $j=\nabla^2_\perp \psi$ is the current density and $\omega=\nabla^2_\perp \phi$ is the vorticity. Note that it would be incorrect to suggest that system (\ref{rmhd2}) describes %%@
quasi-two-dimensional turbulence, since the linear terms describing the field-parallel dynamics, that is, the terms containing $\partial_z$,  are of the same order as the nonlinear terms.

The system (\ref{rmhd2}) is  the best known form of the RMHD equations. The less common symmetric Elsasser form (\ref{rmhd}), on the other hand, has advantages for analytic study, for example, it allows %%@
one to derive relations analogous to those by Politano \& Pouquet (1998) for the case of anisotropic MHD turbulence (see \cite{perez_b08}). Both full and RMHD systems can be used for %%@
numerical simulations of strong MHD turbulence, however, the reduced MHD system has half as many independent variables as the full MHD equations, allowing one to speed up the numerical %%@
computations by a factor of two.  

Recently, it has been demonstrated that the system (\ref{rmhd}) can also be effectively 
used for numerical simulations of the universal regime of {\em weak} MHD turbulence (Perez \& Boldyrev 2008). This is somewhat surprising and requires an explanation. As we discussed in %%@
Section~\ref{weak_turbulence}, the weak turbulence spectrum becomes  universal at $k_\perp \gg k_\|$, when the pseudo-Alfv\'en modes decouple from the cascade dynamics. Simultaneously, one needs to satisfy the %%@
condition of weak turbulence (\ref{weak-turb}). In order to satisfy
both conditions, one needs quite a low ratio of $b/B_0$, which implies quite a short Alfv\'en time  and an increased  %%@
computational cost. It turn out that the first condition can be relaxed if one uses the system  where the pseudo-Alfv\'en modes are explicitly removed. This system is obtained from the full MHD system (\ref{mhd-elsasser}) if one sets $z^\pm_\|\equiv 0$. The resulting restricted system then formally coincides with (\ref{rmhd}), with the exception that it now can be used out of the region of validity of the RMHD equations, that is, $k_\perp \gg k_\|$.  The restricted system explicitly neglects the pseudo-Alfv\'en modes, and, %%@
therefore, it describes the universal regime of weak MHD turbulence of shear-Alfv\'en waves as long as the driving force ensures the weak turbulence condition~(\ref{weak-turb}). The condition~$k_\perp \gg %%@
k_\|$ is therefore not required. 

To support this argument, one can demonstrate that the weak turbulence derivation based on the reduced system (\ref{rmhd}) leads to exactly the same kinetic equations~(\ref{galtier-eq}), as %%@
the full MHD system~(\cite{galtier_nnp02,galtier_c06}, Perez \& Boldyrev 2008, Galtier 2009). Moreover, direct numerical simulations of system~(\ref{rmhd}) with the broad $k_\|$-spectrum of the driving force, %%@
necessary to ensure the weak turbulence condition~(\ref{weak-turb}), reproduce the energy spectrum of weak turbulence~$k_\perp^{-2}$ (Perez \& Boldyrev 2008), see Figure~(\ref{balanced_weak}). To date, this has been the only %%@
available direct numerical study of  the universal regime of weak MHD turbulence.  We conclude that the reduced MHD system provides an effective framework for simulating the universal %%@
regimes of MHD turbulence, providing the forcing is chosen accordingly. Depending on the $k_\|$ spectral width of the forcing, the driven turbulence is weak if the fluctuations satisfy the %%@
inequality~(\ref{weak-turb}), and it is strong if approximate equality
holds in~(\ref{weak-turb}) instead. Further examples of numerical
simulations of various regimes of MHD turbulence in the RMHD framework
are available (e.g.,  \cite{oughton_etal04,rappazzo_etal2007,rappazzo_etal2008}, Dmitruk {\em et~al.} 2003). 

It should be reiterated that even if the turbulence is driven in a ``weak'' fashion, it remains weak only for a certain range of scales, and it becomes strong below the field-perpendicular %%@
scales at which the condition (\ref{weak-turb}) breaks down.   With rapidly increasing capabilities of numerical simulations it should become possible to reproduce a large enough inertial %%@
interval, and to observe the predicted transition from weak to strong MHD turbulence.

\subsection{Unbalanced turbulence}
In the preceding sections we assumed that the MHD turbulence was statistically balanced, that is, the energies of counter-propagating Elsasser modes $E^+$ and $E^-$ were the same. This, %%@
however, is by no means guaranteed, since the Elsasser energies are independently conserved by the ideal MHD equations. In nature and in the laboratory (e.g., the solar wind, interstellar %%@
medium, or fusion devices) MHD turbulence is typically unbalanced as it is generated by spatially localized sources or instabilities, so that the energy has a preferred direction of %%@
propagation.  Moreover, there is a reason to believe that imbalance is an inherent fundamental property of MHD turbulence. As argued earlier, numerical simulations and phenomenological %%@
arguments indicate that even when the turbulence is balanced overall, it is unbalanced locally, creating patches of positive and negative cross-helicity (e.g., \cite{matthaeus08,perez_b09,boldyrev_mc09}). We have already encountered %%@
this phenomenon when discussing dynamic alignment in Section~\ref{strong_turbulence}. Unbalanced MHD turbulence is also called ``cross-helical,'' as the normalized cross helicity, %%@
$H^C/E=\frac{1}{2}(E^+-E^-)/(E^++E^-)$, is a natural measure of the imbalance. 

In unbalanced MHD turbulence the energies of counter-propagating Elsasser modes are not equal, and, {\em a priori} might not have same scalings. This raises an interesting question as to whether %%@
the MHD turbulence is universal and scale-invariant. Indeed, if the energy spectra in the unbalanced domains are different, the overall energy spectrum does not have to be scale-invariant %%@
and universal, but rather may depend on the driving and dissipation mechanisms. Below we consider the cases of strong and weak unbalanced turbulence separately. \\

{\em Strong unbalanced turbulence}. Phenomenological treatments of strong unbalanced MHD turbulence are complicated by the fact one can formally construct two time scales for nonlinear energy %%@
transfer. The MHD system (\ref{mhd-elsasser}) suggests that the times of nonlinear deformation of the $z^\pm$ packets are $\tau^\pm\sim \lambda/z_\lambda^\mp$. These time scales are %%@
essentially different in the unbalanced case, which is hard to reconcile with the assumption that most effectively interacting counter-propagating eddies have comparable field-parallel and %%@
field-perpendicular dimensions. Several  phenomenological models attempting to accommodate this time difference have been proposed recently. These, however, have led to conflicting %%@
predictions for the turbulent spectra of~$E^+$ and~$E^-$ (\cite{lithwick_gs07,beresnyak_l08,chandran08,perez_b09}).      

A possible resolution has been proposed (\cite{perez_b09}). Here the phenomenon of scale-dependent dynamic alignment was invoked to estimate the interaction times.  In an unbalanced eddy, %%@
the field-perpendicular fluctuations of ${\bf v}_\lambda$ and ${\bf b}_\lambda$ are aligned in the field-perpendicular plane within a small angle $\theta_\lambda$. In the phenomenology of %%@
scale-dependent dynamic alignment, these  fluctuations are almost normal to the direction of their gradient, which is also in the field-perpendicular plane. In the case of strong imbalance, %%@
$z^+\gg z^-$, ${\bf z}^+$ and ${\bf z}^-$ then form different angles with the direction of the gradient, and a geometric constraint requires that $z^+_\lambda\theta^+_\lambda \sim %%@
z^-_\lambda\theta^-_\lambda$. In the aligned case the nonlinear interaction time is increased by the alignment angle, $\tau^\pm\sim \lambda/z^{\mp}_\lambda\theta_\lambda^\mp$, leading to the %%@
conclusion that the nonlinear interaction times are the {\em same} for both $z^+$ and $z^-$ packets. Therefore, $z_\lambda^+$ has the same scaling as $z_\lambda^-$, and $\theta_\lambda^+$ has the same scaling as $\theta_\lambda^-$.  

As a result, in strong unbalanced MHD turbulence the spectra of the two sets of modes should have the same scaling, $E^+(k_\perp)\sim E^-(k_\perp)\propto k_\perp^{-3/2}$, but different %%@
amplitudes. This result is consistent with the picture that overall balanced MHD turbulence consists of regions of local imbalance of various strengths. In each of the unbalanced regions the %%@
fluctuations are dynamically aligned and the discussed phenomenology applies. The scaling of the spectrum of strong MHD turbulence is therefore universal, and it does not depend on the %%@
degree of overall imbalance of the turbulence. These results seem to be supported by numerical simulations (\cite{perez_b09,perez_b10}). 

\begin{figure} 
\centering
\includegraphics[width=1.0\textwidth]{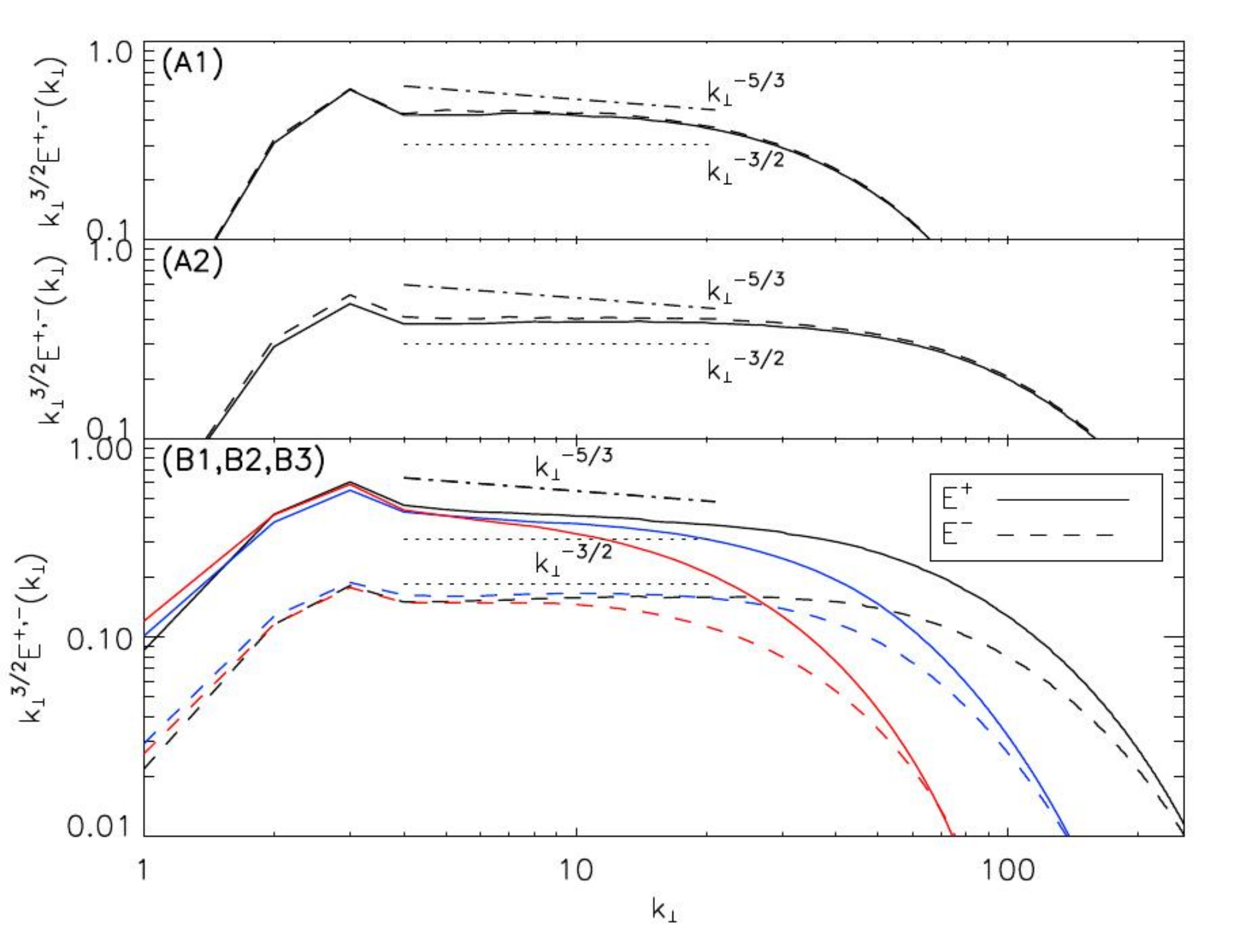}
%\centerline{\psfig{file=strong-balanced-imbalanced.eps,width=5.2in,angle=0}}
\caption{Compensated field-perpendicular spectra of strong balanced
  MHD turbulence (A1-A2), and strong unbalanced MHD turbulence (B1-B3)
  in numerical simulations of reduced MHD system. Runs A1, A2 have
  Reynolds numbers $Re=Rm=2400$ and $6000$, correspondingly. Runs B1,
  B2, and B3 have Reynolds numbers $900$, $2200$, and $5600$. In runs
  B1-B3, the slopes of the Elsasser spectra become progressively
  more parallel and close to -3/2 as the Reynolds number increases (from \cite{perez_b10}).}  
\label{strong-balanced-imbalanced}
\end{figure}

\begin{figure} 
\centering
\includegraphics[width=1.0\textwidth]{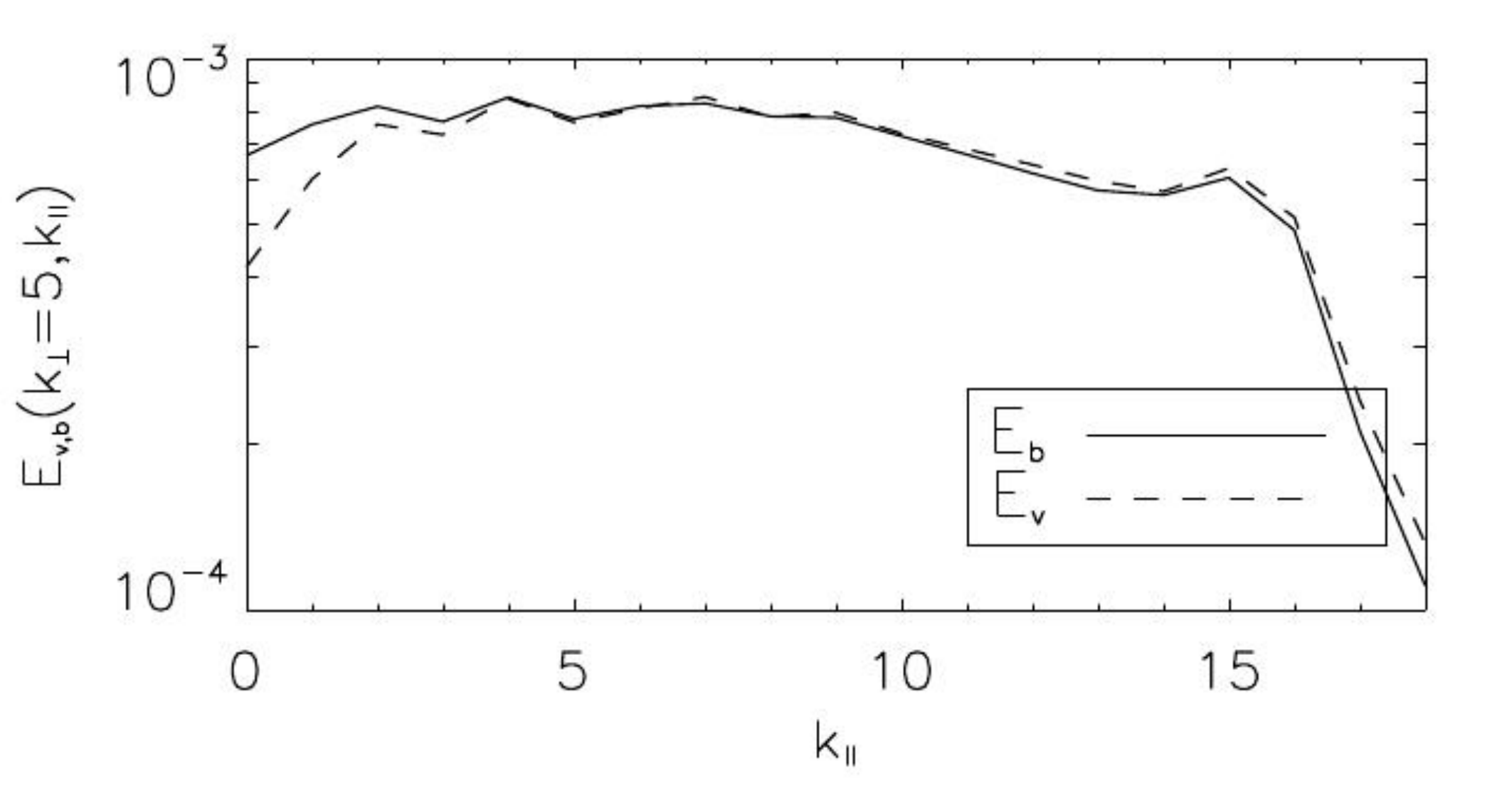}
%\centerline{\hskip-5mm\psfig{file=evb_kpar.eps,width=5.2in,angle=0}}
\caption{Field-parallel magnetic and kinetic energy spectra at given $k_\perp=5$ in the inertial interval. The condensate of the residual energy appears at small $k_\|$. The simulations correspond to moderately imbalanced weak MHD turbulence $z^+/z^-\sim 2$, resolution $512^2\times 256$, Reynolds number $Re=Rm=2500$ (from \cite{boldyrev_p09}).}  
\label{evb_kpar}
\end{figure}

\begin{figure} 
\centering
\hskip-5mm
\includegraphics[width=1.0\textwidth]{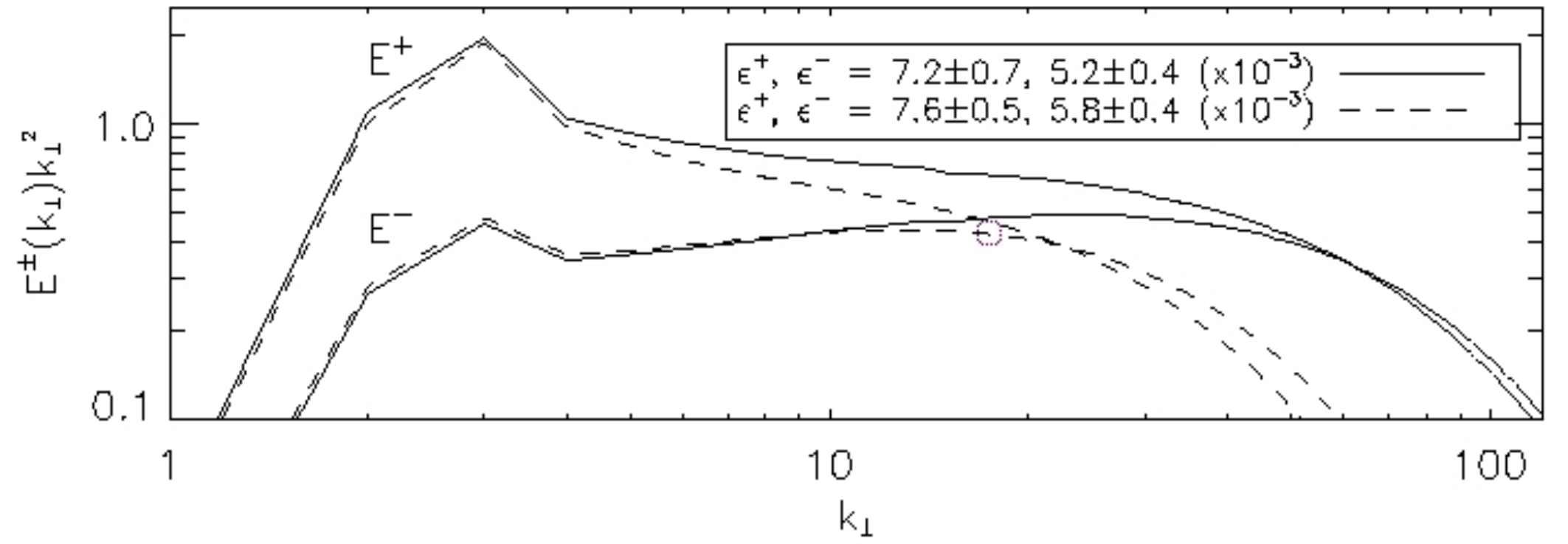}
%\centerline{\hskip-5mm\psfig{file=pinning.eps,width=5.in,angle=0}}
\caption{Compensated field-perpendicular spectra of the Elsasser
  fields in unbalanced weak MHD turbulence simulations based on
  reduced MHD equations.  Resolution $1024^2\times 256$, Reynolds
  numbers $Re=Rm=2000$ and $4500$. The spectra are anchored at large
  scales and pinned at the dissipation scales. As the Reynolds number
  increases, the spectral slopes should become progressively more
  parallel thus converging to~-2 (from \cite{boldyrev_p09}).}  
\label{pinning}
\end{figure}

To conclude this section we note that the above consideration allows one to predict the scalings of the Elsasser fields, however, it does not allow one to specify their amplitudes. Since the $E^+$ energy is mostly concentrated in positively aligned eddies, while $E^-$ energy in negatively aligned ones, the amplitudes of the $E^{\pm}$ spectra averaged over the whole turbulent domain depend on the numbers of eddies of each kind.  In particular, for each eddy one can estimate $(z^+_\lambda /z^-_\lambda)^2 \sim \epsilon^+ / \epsilon^-$, however, this relation should not generally hold for the quantities $\langle (z^{\pm}_\lambda)^2 \rangle $ and $\langle \epsilon^{\pm} \rangle $ averaged over the whole turbulent domain (e.g., \cite{perez_b10b}). In principle, such an average may be not  universal, but rather may depend on the structure of the large-scale driving. 
A possible refinement of the theory, which takes into account different populations of positively and negatively aligned eddies has been proposed (\cite{podesta_b09}).\\

{\em Weak unbalanced turbulence}. Weak unbalanced MHD turbulence allows for a more detailed consideration. A good starting point is provided by the equations~(\ref{galtier-eq}) describing %%@
the evolution of the Elsasser energies (\cite{galtier_nnp00}). We recall that in deriving these equations one assumes that in the zeroth-order approximation, only the %%@
correlation functions $e^{\pm}({\bf k}) \propto \langle {\bf z}^\pm ({\bf k})  \cdot {\bf  z}^\pm (-{\bf k}) \rangle $ are non-zero. So far we have discussed the ``balanced'' solution of this system, 
$e^+({\bf k})= e^-({\bf k}) \propto k_\perp^{-3}$. The authors realized that the system can also describe unbalanced MHD turbulence.  They noticed that  the system %%@
has a broader range of steady solutions: the right hand side integrals of equation~(\ref{galtier-eq}) vanish for a one-parameter family of power-like solutions   
\begin{eqnarray}
e^{\pm}({\bf k})=g^{\pm}(k_\|)k_{\perp}^{-3\pm \alpha},
\label{spectra}
\end{eqnarray} 
with arbitrary functions $g^{\pm}(k_\|)$, which are smooth at $k_\|= 0$, and $-1<\alpha<1$. For $\alpha \neq 0$ these solutions correspond to unbalanced MHD turbulence.  The different energy spectra correspond to different %%@
energy fluxes over scales, which in a steady state are equal to the rates of energy
provided by the large-scale forcing to the Elsasser fields; we denote these rates as~$\epsilon^{\pm}$. One can demonstrate directly from eq.~(\ref{galtier-eq}) that $\alpha$ is uniquely %%@
determined if the ratio of the fluxes $\epsilon^+/\epsilon^-$ is specified. One can further show that the solution with the steeper spectrum corresponds to the larger energy flux, and vice %%@
versa~(\cite{galtier_nnp00,lithwick_g03}).

Once the energy fluxes provided by the large-scale forcing are
specified, the slopes of the energy spectra are specified but their
amplitudes are not. Fully to remove the degeneracy, 
one further argues that at the dissipation scale the balance should be
restored, that is,  $e^+(k)$ should converge to $e^-(k)$. This
so-called ``pinning'' effect was first pointed out in  Grappin {\it et
  al} (1983), and has been elaborated upon in greater detail (see
e.g. \cite{galtier_nnp00,lithwick_g03,chandran08}).  The pinning can be physically understood, if one notes that the $e^+(k)$ and $e^-(k)$ energy %%@
spectra cannot intersect in the inertial interval, as this would contradict the universality of the turbulence. The alignment of velocity and magnetic fluctuations, preserved by the nonlinear %%@
terms, can be broken only by the dissipation. This pinning effect is indeed observed in the simulations presented below.  

According to the above picture, if the rates of energy supply are fixed, then the {\em slopes}  of the energy spectra $e^{\pm}(k)$ are fixed as well. If the dissipation scale is now changed, %%@
the {\em amplitudes} of the spectra should change so as to maintain the specified slopes and to make them converge at the dissipation scale. This conclusion, although consistent with %%@
equations (\ref{galtier-eq}), seems however to be at odds with common intuition about turbulent systems, which suggests that small-scale dissipation should not significantly affect %%@
large-scale fields subject to the same large-scale driving. 

This seeming contradiction has recently been addressed (\cite{boldyrev_p09}). It was  proposed that driven weak MHD turbulence generates a residual energy condensate  
\begin{eqnarray}
\langle {\bf z}^+({\bf %%@
k})\cdot {\bf z}^{-}(-{\bf k})\rangle = v^2-b^2 \neq 0 \quad \mbox{at} \,\, k_\|=0.
\label{condensate}
\end{eqnarray}
This condensate has been assumed to be zero in the standard
derivations (Ng \& Bhattacharjee 1996,\cite{goldreich_s97,galtier_nnp00}), %%@
however, this assumption is not necessary and apparently incorrect in the unbalanced case.  The presence of the condensate can be explained as follows. Alfv\'en wave fluctuations at $k_\| %%@
\neq 0$ obey ${\bf v}= \pm {\bf b}$, in which case the residual energy vanishes. However, at $k_\|=0$ fluctuations are not waves, and the  Alfv\'enic condition should not be necessarily %%@
satisfied. The presence of the condensate (\ref{condensate}) implies that  magnetic and kinetic energies are not in equipartition at $k_\|=0$. 

It can be argued that the generation of the %%@
condensate is a consequence of the breakdown of the mirror symmetry in unbalanced turbulence. Indeed, non-balanced MHD turbulence is not mirror-invariant, as it possesses  non-zero %%@
cross-helicity, i.e. $H^c=\int ({\bf v}\cdot {\bf b})d^3x=(E^+-E^-)/4 \neq 0$. Non-mirror invariant turbulence can also possess nonzero magnetic helicity, $H^m=\int ({\bf A}\cdot {\bf B})\, d^3x \neq 0$. Helical magnetic fields should not necessarily be in equipartition with the velocity field, as is obvious from the fact that the corresponding Lorentz force may be equal to zero (consider, for example, a Beltrami field ${\bf B}=\alpha \nabla \times {\bf B}$, which is maximally helical, but exerts no force on the velocity field). In the case of weak MHD turbulence, the only region of phase space where the equipartition between the magnetic and velocity fluctuations may be broken is the region $k_\|=0$, where such fluctuations are not Alfv\'en waves.  In the case of a uniform guide field ${\bf B}_0$, which is relevant for the numerical set-up, these arguments can have further support if one notes that the magnetic helicity of {\em fluctuations} is not conserved, rather, it is generated by the term $\int [{\bf z}^+\times {\bf z}^-]_\|\,d^3 x$, (see e.g., \cite{galtier_nnp00}). Interestingly, the magnetic field of the condensate is generated by the $k_\|=0$ component of the same term, $[{\bf z}^+\times {\bf z}^-]_\|$. Such a condensate is indeed observed in the numerics, see Figure~(\ref{evb_kpar}).

%Non-mirror-invariant turbulence can generate large-scale magnetic fields that are not in equipartition with the velocity field. %%@
%This probably happens owing to conservation of magnetic helicity in ideal MHD, which tends to cascade toward large scales in a turbulent state. Apparently, such a process is preserved in %%@
%driven unbalanced weak MHD turbulence, although in a peculiar fashion --  it leads to generation of  condensate in the vicinity of $k_\|=0$, where the magnetic energy exceeds the kinetic %%@
%energy. Such a condensate is indeed observed in the numerics, see Figure~(\ref{evb_kpar}).
%\begin{figure}
%  \includegraphics[width=0.48\textwidth]{evb_kpar.eps}  
%  \caption{The field-parallel spectra of the magnetic energy (solid line) and the kinetic energy (dashed line) at the field-perpendicular wave number~$k_{\perp}=5$; $Re=2500$, resolution %%@
%$512^2\times 256$ points.}
%  \label{evb_kpar}
%\end{figure}

The conclusion of this analysis are the following. When the turbulence is balanced, the energy spectra are   $E^{\pm}(k_{\perp})\propto k_{\perp}^{-2}$, in agreement with the analytic %%@
prediction   (Ng \& Bhattacharjee 1996, \cite{goldreich_s97,galtier_nnp00}). In the balanced case the evolution of $E^{\pm}$ fields is not affected by the condensate. In the non-balanced case the %%@
interaction with the condensate becomes essential, and the universal power-law spectra may not exist in an inertial interval of limited extent. Both spectra $E^{\pm}(k_{\perp})$ %%@
have the large-scale amplitudes fully specified by the external forcing, and they converge at the dissipation  scale. As the dissipation scale decreases, however, the spectral scalings (but %%@
not amplitudes) approach each other at large~$k_{\perp}$ such that the universal scaling $k_{\perp}^{-2}$ is recovered for both the $E^{\pm}(k_{\perp})$ spectra as  $k_{\perp}\to \infty$, see Figure~(\ref{pinning}).

Finally, we would like to point out that unbalanced MHD turbulence is  numerically more challenging than balanced turbulence. Indeed, in the case $z^+ \gg z^-$ 
the formal Reynolds number associated with the $z^+$ mode is much smaller than the Reynolds number associated with $z^-$. The stronger the imbalance the larger 
the Reynolds number and the resolution required to reproduce the inertial intervals for both fields. For example,  numerical simulations with the resolution of $1024^2$ points in the field-perpendicular direction %%@
do not allow one to address turbulence with imbalance stronger than $z^+/z^-\sim 2$--$3$, (see e.g., \cite{perez_b10}).

%% file: conclusions.tex
\section{Conclusions}

The last two decades have seen many developments in the field of MHD turbulence. There is little doubt that the driving force behind many of these has been the remarkable %%@
increase in computing power. In a field in which it so difficult to carry out experiments or detailed observations, researchers have relied almost entirely on numerical simulations for the %%@
underpinning of their theoretical speculations and modelling. It is now a matter of routine to carry out simulations with in excess of one billion grid-points; and the expectation is that in %%@
the near future simulations with $10^{12}$ grid-points will become feasible. With these resolutions it is possible to explore fully three-dimensional configurations with moderately high %%@
Reynolds numbers. In other words, the turbulent regime has become available to computations. Given these advances and these possibilities, it is natural to conclude this review with some %%@
assessment of what has been learned --- and is now considered fairly certain ---  and what still remains unclear or controversial, and will probably occupy the minds of researchers, and the CPU's of %%@
their computers, in the years to come. 

Within the framework of dynamo theory it is now widely accepted that given a random velocity, dynamo action is always possible provided the magnetic Reynolds number is large enough. In the %%@
case of reflectionally symmetric turbulence (i.e.\ non-helical)  the effort required to drive a dynamo appears to be determined by the slope of the energy spectrum of the velocity at %%@
spatial scales comparable to those on which magnetic reconnection occurs. The flatter the spectrum the harder it is to drive a dynamo. If this property is expressed in terms of the %%@
velocity roughness exponent, a marked difference emerges between the mechanisms leading to dynamo action for a smooth velocity as opposed to those in which the velocity is rough. This %%@
distinction between smooth and rough velocities then maps very naturally in two distinct regimes corresponding to large and small values of the magnetic Prandtl number. What is also clear now, is that this stochastic theory must be modified if coherent structures are present in the turbulence. These may, depending on their properties, take over the control of the dynamo growth rate. We are now only beginning to %%@
understand which properties of coherent structures contribute to the dynamo process, and certainly this will be an active area of future research. In the nonlinear regime, there are now %%@
compelling models for the saturation process leading to the establishment of a stationary turbulent state for dynamos at large Magnetic Prandtl number. These models, in general, rely on some %%@
specific geometrical property of the the magnetic field, like foliation, and seem to be borne out by the existing simulations. It remains to be seen if the predictions of these, mostly %%@
phenomenological models, continue to hold at higher values of $Rm$. By contrast, the low magnetic Prandtl number regime remains largely unexplored. The difficulties are both conceptual, %%@
since the velocity correlator has a non-analytic behaviour at the reconnection scales, and numerical, since reproducing a wide separation of scales between the magnetic and velocity boundary %%@
layers is extremely computationally expensive; in this review we have determined the size of the calculation required to begin to settle the issue. The general expectation is, however, that the turbulence should become independent of of the magnetic Prandtl number provided the latter is small enough. %%@
The existence of this asymptotic regime remains conjectural pending the availability of either much bigger computers, or much better analytical approaches. 

When the flows are affected by the large scale component of the magnetic field, irrespective of whether the latter is self generated or imposed externally, the techniques utilised in analysing the turbulence are different.
Two  important ideas have emerged that %%@
have strongly influenced our understanding of the turbulent process in this case. The first is that depending on the dominant mechanism responsible for energy transfer MHD turbulence can be either weak %%@
or strong. In the weak case, mean-fluctuation interactions dominate, whereas in the strong case, mean-fluctuation and fluctuation-fluctuation interactions are comparable. The common wisdom is %%@
that the strength of the fluctuation-fluctuation interactions increases at small scales, so that if the inertial range is sufficiently extended there will always be a transition from weak to %%@
strong turbulence. At present there are no cases in which this transition has been convincingly demonstrated numerically. The problem is associated with difficulties in reproducing a deep %%@
inertial range. The situation will no doubt improve with the next generation of supercomputers. The second important idea is that MHD turbulence, unlike its unmagnetised counterpart, has a richly geometrical %%@
structure. There is now universal agreement that the energy cascade is anisotropic with most of the energy being transferred transverse to the (local) mean field. The exact geometry of this %%@
process is controlled by the requirement of a ``critical balance" between the crossing time of counter propagating wavepackets and the time for nonlinear transfer in the %%@
field-perpendicular direction. Furthermore, there are strong indications, both analytical and numerical, that the polarization vectors of magnetic and velocity fluctuations tend to align in %%@
the transverse plane. Although this  phenomenon is intimately connected with the forward cascade of cross-helicity it has measurable consequences for the slope of the energy spectrum. This %%@
appears shallower than predicted in the absence of alignment. Both the elongation of the wavepackets and the degree of alignment increase at small scales, and hence the picture that emerges of MHD %%@
turbulence in the deep inertial range is that of a collection of interacting ribbon-like structures. Another problem that has recently emerged, and is at the moment the focus of considerable %%@
interest, is that of unbalanced cascades. This occurs when there is an excess of wavepackets propagating in one direction relative to those propagating in the opposite %%@
direction. The basic question here is whether unbalanced MHD turbulence is fundamentally different from balanced turbulence, or whether both are the same, with  even balanced turbulence %%@
being made up of interwoven unbalanced patches that appear balanced only on average. Again, a definite resolution of these issues may have to wait for a further increase in computing resources.

Finally, we should remark about the areas of active research in MHD turbulence that are not discussed in this review. In dynamo theory, the most important problem remains that of the %%@
generation of large scale fields by turbulence lacking reflectional symmetry. At the moment, there appears to be a fracture between the conventional wisdom, mostly grounded in mean-field %%@
theory, and an increasing body of numerical experiments. It is
possible that  large-scale dynamos are ``essentially nonlinear" rather
than ``essentially kinematic" and a major revision of the theory might
be in order. It is also possible that the whole categorisation of dynamos into large and small-scale
is misleading and that it is more useful to seek ``system-scale'' dynamo solutions to the MHD equations (\cite{tcb:2010}).
In general, the lack of reflectional symmetry in
astrophysical turbulence owes its origins to the presence of
rotation. Strong rotation and shear flows are known to modify the
nature of MHD turbulence and indeed may lead to the generation of new
instabilties (\cite{chan:61,balhaw:1991}). Such magnetorotational
turbulence is of great interest owing to its importance for the
accretion process, and the nature of the turbulence may be very
different to that discussed in this review (see
e.g. \cite{balhaw:1998} and the references therein).

For relectionally symmetric turbulence, the most natural extension of %%@
what has been presented is to incorporate compressibility. This bring in one more set of linear waves --- the fast magnetosonic waves --- and modifies the nature of the pseudo-Alfv\'en waves. The %%@
resulting turbulence is extremely rich and is only now beginning to be explored in the nonlinear regime (see e.g. \cite{berlazcho:2005,lietal:2008,chandran08b}). Finally, we should note that all of our considerations have been discussed within the %%@
framework of classical MHD. There is an almost bewildering variety of new effects that become important when the particular nature of a plasma is taken into account (see e.g. \cite{kulsrud:2005}). Again, the full richness %%@
of this system will be a major preoccupation in the near future.